\documentclass[aip,jmp,reprint,a4paper,amsmath,onecolumn,nofootinbib]{revtex4-1}
\usepackage[utf8]{inputenc}
\usepackage[colorlinks=true,urlcolor=blue,anchorcolor=blue,citecolor=blue,filecolor=blue,linkcolor=blue,menucolor=blue,linktocpage=true,pdfa=true,unicode]{hyperref}
\usepackage{bm}
\usepackage{lmodern}

\newcommand{\eqend}[1]{\,\mathrm{#1}}
\newcommand{\bra}[1]{{\left\langle{#1}\right\vert}}
\newcommand{\ket}[1]{{\left\vert{#1}\right\rangle}}
\newcommand{\abs}[1]{{\left\vert{#1}\right\vert}}
\newcommand{\laplace}{\bigtriangleup}
\newcommand{\hankel}[4][]{\mathrm{H}^{(#2)}_{#3}#1\left(#4\right)}
\newcommand{\besselj}[2]{\mathrm{J}_{#1}\left(#2\right)}

\newcommand{\besselk}[2]{\mathrm{K}_{#1}\left(#2\right)}
\newcommand{\appell}[1]{\mathrm{F}_#1}
\newcommand{\hypergeom}[2]{\,{}_{#1}\mathrm{F}_{#2}}
\newcommand{\sgn}{\mathop{\mathrm{sgn}}}
\newcommand{\bigo}[1]{\mathcal{O}\left({#1}\right)}
\newcommand{\total}{\mathop{}\!\mathrm{d}}
\renewcommand{\vec}[1]{\bm{#1}}
\newcommand{\mathi}{\mathop{}\!\mathrm{i}}
\newcommand{\mathe}{\mathop{}\!\mathrm{e}}

\begin{document}

\title{Mode-sum construction of the two-point functions for the Stueckelberg vector fields in the Poincaré patch of de~Sitter space}

\author{Markus B. Fröb}
\email{mfroeb@ffn.ub.edu}
\affiliation{Departament de Física Fonamental, Institut de Ciències del Cosmos (ICC), Universitat de Barcelona (UB), C/ Martí i Franquès 1, 08028 Barcelona, Spain}
\affiliation{Department of Mathematics, University of York, Heslington, York, YO10 5DD, United Kingdom}
\author{Atsushi Higuchi}
\email{atsushi.higuchi@york.ac.uk}
\affiliation{Department of Mathematics, University of York, Heslington, York, YO10 5DD, United Kingdom}

\date{11.\,Dec.\,2013, revised 17.\,Feb.\,2014 and 06.\,May\,2014}

\begin{abstract}
We perform canonical quantization of the Stueckelberg Lagrangian for massive vector fields in the conformally flat patch of de~Sitter space in the Bunch-Davies vacuum and find their Wightman two-point functions by the mode-sum method. We discuss the zero-mass limit of these two-point functions and their limits where the Stueckelberg parameter $\xi$ tends to zero or infinity. It is shown that our results reproduce the standard flat-space propagator in the appropriate limit. We also point out that the classic work of Allen and Jacobson for the two-point function of the Proca field and a recent work by Tsamis and Woodard for that of the transverse vector field are two limits of our two-point function, one for $\xi\to \infty$ and the other for $\xi\to 0$. Thus, these two works are consistent with each other, contrary to the claim by the latter authors.
\end{abstract}

\pacs{04.62.+v, 04.60.Ds, 11.15.-q, 14.70.-e}

\maketitle

\section{Introduction}

Massive vector fields arise from spontaneous symmetry breaking in the electroweak sector of the standard model of particle physics. While most inflationary models are based on scalar fields, massive vectors should also play a role and it is important to check if they can give rise to observable effects. To calculate quantum corrections to observables, a fundamental quantity is the propagator of such a field, which has been derived for maximally symmetric spaces by Allen and Jacobson~\cite{allenjacobson1986}. They used the equation of motion for the Proca theory~\cite{proca1936} together with the assumption of invariance under the symmetries of the background spaces, and required the singularity structure in the coincidence limit to be the one inferred from the flat-space case. Among the maximally symmetric spaces, de~Sitter space is of particular relevance to cosmology since the geometry of the universe in its inflationary phase is approximately de~Sitter.

However, the validity of this classic result was questioned recently by Tsamis and Woodard~\cite{tsamiswoodard2007}, who claimed that the result of Allen and Jacobson needed to be amended by an additional term. In this paper we reconcile this apparent disagreement by performing a canonical quantization of an extension of the Proca theory proposed by Stueckelberg~\cite{stueckelberg1938} and finding the corresponding two-point functions using the mode-sum method in the conformally flat coordinate patch, which is often called the Poincaré patch. In the limit where the Stueckelberg parameter $\xi$ goes to infinity, the Proca theory is recovered, and in this limit we shall reproduce the result of Allen and Jacobson. We also reproduce the result of Tsamis and Woodard in the limit $\xi\to 0$.
There have been previous works on canonical quantization of massive vector fields in various portions of the de~Sitter space~\cite{spindelschomblond1976,higuchi1987,gazeautakook2000,cotaescu2010}. However, to the best of our knowledge quantization of the general Stueckelberg Lagrangian in $n$ dimensions has not been done.

In the case of a fundamental massive vector field (the original Proca theory), the propagator is well-defined in the free theory, but it is known from flat space that a) such a theory cannot be extended to a consistent interacting theory~\cite{itzyksonzuber} and that only mass terms generated through spontaneous symmetry breaking are viable, and b) that the result is divergent as $m \to 0$. However, the origin of the mass term in the Lagrangian is only important for interactions, and immaterial for the construction of the propagator on which we focus in this work. Nevertheless, to obtain a result which is useful also as $m \to 0$, we add the standard covariant gauge-fixing term to the action which is needed in the case of spontaneous symmetry breaking. The total action including the mass and gauge-fixing terms corresponds exactly to the Stueckelberg theory.

This article is structured as follows: first we perform the canonical quantization of the Stueckelberg Lagrangian in full generality in Sec.~\ref{quantization}. In particular, we find the Wightman two-point function of a massive vector field for general Stueckelberg parameter. In Sec.~\ref{limits} we discuss the limits of zero mass and zero and infinite Stueckelberg parameter of this two-point function. In particular, we point out that the limits $m^2 \to 0$ and $\xi \to \infty$ lead to divergences when they are both taken, as is well known from flat space~\cite{itzyksonzuber}.
In Sec.~\ref{flatspacelimit} it is demonstrated that our two-point function has the correct flat space limit, and in section~\ref{divergencefeynman} we calculate the divergence of the Feynman propagator which generally differs from the divergence of the Wightman two-point function.
We point out that the wrong assumption about the divergence of the Proca field propagator made by Tsamis and Woodard~\cite{tsamiswoodard2007} is the cause of the apparent disagreement of their result with the classic one of Allen and Jacobson~\cite{allenjacobson1986}. We further show that this assumption, however, is satisfied in the Landau gauge $\xi \to 0$, which corresponds to imposing the gauge condition $\nabla_s A^s = 0$ by a functional delta in the path integral.
In Sec.~\ref{comparison} the relation of our results to the one by Allen and Jacobson as well as that by Tsamis and Woodard is clarified. Some technicalities are delegated to the appendices; in Appendix~\ref{appA} we present the evaluation of an integral which is essentially the mode-sum calculation of the scalar-field Wightman function in the Poincaré patch of $n$-dimensional de~Sitter space. In Appendix~\ref{appB} we present details of the mode-sum calculation of the space-space component of our Wightman function, and finally in Appendix~\ref{appC} the massless limits of our two-point function in dimensions $2$, $3$ and $4$ are presented.

We use the convention in which the metric is mostly plus and $R_{ab} = (n-1) H^2 g_{ab}$ for de~Sitter space, where $n$ is the spacetime dimensionality. All formulas for special functions were taken from Ref.~\onlinecite{dlmf}.

\section{Canonical quantization of the Stueckelberg Lagrangian} \label{quantization}

The Stueckelberg Lagrangian for a massive vector field with mass $m$ and Stueckelberg parameter $\xi$ reads
\begin{equation}
\label{st_lag}
\mathcal{L} = - \frac{1}{4} \sqrt{-g} \left( F^{ab} F_{ab} + 2 m^2 A^b A_b + \frac{2}{\xi} \left( \nabla_b A^b \right)^2 \right) \eqend{.}
\end{equation}
In the conformally flat coordinate system of the Poincaré patch
\begin{equation}
\total s^2 = \frac{1}{(-H\eta)^2} \left( - \total \eta^2 + \total \vec{x}^2 \right) \eqend{,}
\end{equation}
with $\eta \in (-\infty,0)$ being the conformal time, it takes the form
\begin{equation}
\begin{split}
\mathcal{L} &= - \frac{1}{2 \xi} (-H\eta)^{4-n} \left( - A'_0 + \partial_\alpha A^\alpha + \frac{n-2}{\eta} A_0 \right)^2 - \frac{1}{2} m^2 (-H\eta)^{2-n} ( - A_0^2 + A^\alpha A_\alpha ) \\
&\quad- \frac{1}{2} (-H\eta)^{4-n} \left( - ( A^{\prime \alpha} - \partial^\alpha A_0 ) ( A'_\alpha - \partial_\alpha A_0 ) + F^{\alpha\beta} ( \partial_\alpha A_\beta ) \right) \eqend{.}
\end{split}
\end{equation}
Note that we use Greek letters $\alpha$, $\beta$, $\gamma$ and so on to denote spatial components, and that such indices are raised and lowered with the flat metric $\eta_{ab}$, e.g.\ $A^\alpha = \eta^{\alpha b} A_b = \delta^{\alpha\beta} A_\beta$. Also, a prime denotes the derivative with respect to the conformal time $\eta$.

The canonical momenta $\pi^a = \partial \mathcal{L}/\partial A'_a$ are readily found to be
\begin{subequations}
\begin{align}
\pi^\alpha &= (-H\eta)^{4-n} ( A^{\prime\alpha} - \partial^\alpha A_0 ) \eqend{,} \\
\pi^0 &= \frac{1}{\xi} (-H\eta)^{4-n} \left( - A'_0 + \partial_\beta A^\beta + \frac{n-2}{\eta} A_0 \right) \eqend{.}
\end{align}
\end{subequations}
The equations of motion following from the Stueckelberg Lagrangian~\eqref{st_lag} are
\begin{subequations}
\label{eom_firstorder}
\begin{align}
A'_\alpha &= (-H\eta)^{n-4} \pi_\alpha + \partial_\alpha A_0 \eqend{,} \\
A'_0 &= - (-H\eta)^{n-4} \xi \pi^0 + \partial_\beta A^\beta + \frac{n-2}{\eta} A_0 \eqend{,} \\
\pi^{\alpha\prime} &= \partial^\alpha \pi^0 + (-H\eta)^{4-n} \partial_\beta F^{\beta\alpha} - m^2 (-H\eta)^{2-n} A^\alpha \eqend{,} \\
\pi^{0\prime} &= - \frac{n-2}{\eta} \pi^0 + \partial_\alpha \pi^\alpha + m^2 (-H\eta)^{2-n} A_0 \eqend{.}
\end{align}
\end{subequations}
We may now postulate canonical commutation relations,
\begin{equation}
\label{ccr}
[ A_a(\vec{x}, \eta), \pi^b(\vec{x}', \eta) ] = \mathi \delta_a^b \delta^{n-1}(\vec{x}-\vec{x}') \eqend{.}
\end{equation}
Commutators of the components of the field $A_a$ as well as of those of the canonical momentum $\pi^a$ vanish.

To decouple the field equations, we separate the field and the canonical momentum into the transverse and longitudinal parts,
\begin{equation}
A_\alpha = B_\alpha + \partial_\alpha A \eqend{,} \qquad \pi^\alpha = \varpi^\alpha + \partial^\alpha \pi \eqend{,}
\end{equation}
with $\partial^\alpha B_\alpha = \partial_\alpha \varpi^\alpha = 0$. After some rearrangements, we obtain for the transverse parts
\begin{subequations}
\label{eom_transverse}
\begin{align}
B''_\alpha - \frac{n-4}{\eta} B'_\alpha - \laplace B_\alpha &= - (-H\eta)^{-2} m^2 B_\alpha \eqend{,} \\
\varpi_\alpha &= (-H\eta)^{4-n} B'_\alpha \eqend{,}
\end{align}
\end{subequations}
where $\laplace = \partial^\alpha \partial_\alpha$. The longitudinal and temporal components of the field are still coupled. However, it is possible to obtain a decoupled equation for their conjugate momenta:
\begin{subequations}
\label{eom_longitudinal}
\begin{align}
\pi'' + \frac{n-2}{\eta} \pi' - \laplace \pi &= - (-H\eta)^{-2} m^2 \pi \eqend{,} \\
\pi^{0\prime\prime} + \frac{n-2}{\eta} \pi^{0\prime} - \laplace \pi^0 &= - (-H\eta)^{-2} \left( \xi m^2 - (n-2) H^2 \right) \pi^0 \eqend{.}
\end{align}
\end{subequations}
In terms of those, the longitudinal and temporal components of the field are given by
\begin{subequations}
\label{eom_longA}
\begin{align}
A &= - \frac{(-H\eta)^{n-2}}{m^2} \left( \pi' - \pi^0 \right) \eqend{,} \\
A_0 &= \frac{(-H\eta)^{n-2}}{m^2} \left( \pi^{0\prime} + \frac{n-2}{\eta} \pi^0 - \laplace \pi \right) \eqend{.}
\end{align}
\end{subequations}
We then introduce the following decomposition in modes in (spatial) Fourier space:
\begin{subequations}
\label{mode_decomposition}
\begin{align}
\pi^0(\vec{x},\eta) &= \int \left( a_{(0)}(\vec{p}) g_{(0)}(\vec{p},\eta) \mathe^{\mathi \vec{p} \vec{x}} + a^\dagger_{(0)}(\vec{p}) g^*_{(0)}(\vec{p},\eta) \mathe^{-\mathi \vec{p} \vec{x}} \right) \frac{\total^{n-1} \vec{p}}{(2\pi)^{n-1}} \\
B_\alpha(\vec{x},\eta) &= \int \sum_{k=1}^{n-2} e^{(k)}_\alpha (\vec{p}) \left( a_{(k)}(\vec{p}) f_{(k)}(\vec{p},\eta) \mathe^{\mathi \vec{p} \vec{x}} + a^\dagger_{(k)}(\vec{p}) f^*_{(k)}(\vec{p},\eta) \mathe^{-\mathi \vec{p} \vec{x}} \right) \frac{\total^{n-1} \vec{p}}{(2\pi)^{n-1}} \eqend{,} \\
\pi(\vec{x},\eta) &= \int \left( a_{(n-1)}(\vec{p}) g_{(n-1)}(\vec{p},\eta) \mathe^{\mathi \vec{p} \vec{x}} + a^\dagger_{(n-1)}(\vec{p}) g^*_{(n-1)}(\vec{p},\eta) \mathe^{-\mathi \vec{p} \vec{x}} \right) \frac{\total^{n-1} \vec{p}}{(2\pi)^{n-1}} \eqend{.}
\end{align}
\end{subequations}
We note that the field $B_\alpha(\vec{x},\eta)$ is only present for $n>2$, and has to be set to zero if $n=2$. The Fourier decomposition of $\varpi^\alpha(\vec{x},\eta)$, $A(\vec{x},\eta)$ and $A_0(\vec{x},\eta)$ can readily be found from
\eqref{eom_transverse} and \eqref{eom_longA}. The $e^{(k)}_\alpha (\vec{p})$, $k=1,2,\ldots,n-1$ constitute a set of orthonormal polarization vectors with $\abs{\vec{p}} e^{(n-1)}_\alpha (\vec{p}) = p_\alpha$ and
\begin{equation}
\sum_{k=1}^{n-2} e^{(k)}_\alpha (\vec{p}) e^{(k)}_\beta (\vec{p}) = \eta_{\alpha\beta} - \frac{p_\alpha p_\beta}{\vec{p}^2} \eqend{.}
\end{equation}

The mode functions $g_{(0)}(\vec{p},\eta)$, $f_{(k)}(\vec{p},\eta)$  and $g_{(n-1)}(\vec{p},\eta)$ are determined by the equations of motion, \eqref{eom_transverse} and \eqref{eom_longitudinal}, and their normalization can be determined from the canonical commutation relations~\eqref{ccr} together with the commutation relations,
\begin{equation}
[ a_{(k)}(\vec{p}), a_{(l)}^\dagger(\vec{q}) ] = (2\pi)^{n-1} \delta^{n-1}(\vec{p}-\vec{q}) \eta_{kl} \eqend{,}
\end{equation}
with all other commutators vanishing. This procedure works out to give
\begin{subequations}
\begin{align}
g_{(0)}(\vec{p},\eta) g^{*\prime}_{(0)}(\vec{p},\eta) - g^*_{(0)}(\vec{p},\eta) g'_{(0)}(\vec{p},\eta) &= \mathi m^2 (-H\eta)^{2-n} \eqend{,} \\
f_{(k)}(\vec{p},\eta) f^{*\prime}_{(k)}(\vec{p},\eta) - f^*_{(k)}(\vec{p},\eta) f'_{(k)}(\vec{p},\eta) &= \mathi (-H\eta)^{n-4} \qquad (1 \leq k \leq n-2) \eqend{,} \\
g_{(n-1)}(\vec{p},\eta) g^{*\prime}_{(n-1)}(\vec{p},\eta) - g^*_{(n-1)}(\vec{p},\eta) g'_{(n-1)}(\vec{p},\eta) &= \mathi m^2 \abs{\vec{p}}^{-2} (-H\eta)^{2-n} \eqend{.}
\end{align}
\end{subequations}
Of course, what we describe here parallels exactly the flat space case, with the only difference that, because of the time dependence of the metric, we can only work in spatial Fourier space here while in flat space one usually Fourier transforms the time coordinate as well.

For the Bunch-Davies vacuum $\ket{0}$ which is annihilated by the $a_{(k)}(\vec{p})$, we have up to phase factors
\begin{subequations}
\begin{align}
g_{(0)}(\vec{p},\eta) &= m \sqrt{\frac{\pi}{4 H}} (-H\eta)^\frac{3-n}{2} \mathe^{\mathi\mu \frac{\pi}{2}} \hankel1\mu{-\abs{\vec{p}}\eta} \eqend{,} \\
f_{(k)}(\vec{p},\eta) &= \sqrt{\frac{\pi}{4 H}} (-H\eta)^\frac{n-3}{2} \mathe^{\mathi\nu \frac{\pi}{2}} \hankel1\nu{-\abs{\vec{p}}\eta} \qquad (1 \leq k \leq n-2) \eqend{,} \\
g_{(n-1)}(\vec{p},\eta) &= \frac{m}{\abs{\vec{p}}} \sqrt{\frac{\pi}{4 H}} (-H\eta)^\frac{3-n}{2} \mathe^{\mathi\nu\frac{\pi}{2}} \hankel1\nu{-\abs{\vec{p}}\eta} \eqend{,}
\end{align}
\end{subequations}
where the constants $\mu$ and $\nu$ are defined by
\begin{equation}
\label{def_munu}
\mu = \sqrt{\frac{(n-1)^2}{4} - \xi \frac{m^2}{H^2}} \eqend{,} \qquad \nu = \sqrt{\frac{(n-3)^2}{4} - \frac{m^2}{H^2}} \eqend{.}
\end{equation}
The two-point (Wightman) function is then easily calculated to be
\begin{subequations}
\label{AA}
\begin{align}
\bra{0} A_0(x) A_0(x') \ket{0} &= - \frac{H^{n-2}}{(4\pi)^\frac{n}{2} m^2} \left( \laplace I_\nu(\eta,\eta',\vec{x}-\vec{x}') + \partial_\eta \partial_{\eta'} I_\mu(\eta,\eta',\vec{x}-\vec{x}') \right) \eqend{,} \\
\bra{0} A_0(x) A_{\beta'}(x') \ket{0} &= - \frac{H^{n-2}}{(4\pi)^\frac{n}{2} m^2} (H^2\eta\eta')^{n-2} \partial_{\beta'} \partial_{\eta'} \left( (H^2\eta\eta')^{2-n} I_\nu(\eta,\eta',\vec{x}-\vec{x}') \right) \nonumber \\
&\qquad- \frac{H^{n-2}}{(4\pi)^\frac{n}{2} m^2} \partial_\eta \partial_{\beta'} I_\mu(\eta,\eta',\vec{x}-\vec{x}') \eqend{,} \\
\bra{0} A_\alpha(x) A_0(x') \ket{0} &= - \frac{H^{n-2}}{(4\pi)^\frac{n}{2} m^2} (H^2\eta\eta')^{n-2} \partial_\alpha \partial_\eta \left( (H^2\eta\eta')^{2-n} I_\nu(\eta,\eta',\vec{x}-\vec{x}') \right) \nonumber \\
&\qquad- \frac{H^{n-2}}{(4\pi)^\frac{n}{2} m^2} \partial_\alpha \partial_{\eta'} I_\mu(\eta,\eta',\vec{x}-\vec{x}') \eqend{,} \\
\bra{0} A_\alpha(x) A_{\beta'}(x') \ket{0} &= \frac{H^{n-2}}{(4\pi)^\frac{n}{2}} (H^2\eta\eta')^{-1} \left( \eta_{\alpha\beta'} + \frac{\partial_\alpha \partial_{\beta'}}{\laplace} \right) I_\nu(\eta,\eta',\vec{x}-\vec{x}') \nonumber \\
&\qquad- \frac{H^{n-2}}{(4\pi)^\frac{n}{2} m^2} (H^2\eta\eta')^{n-2} \frac{\partial_\alpha \partial_{\beta'}}{\laplace} \partial_\eta \partial_{\eta'} \left( (H^2\eta\eta')^{2-n} I_\nu(\eta,\eta',\vec{x}-\vec{x}') \right) \nonumber \\
&\qquad- \frac{H^{n-2}}{(4\pi)^\frac{n}{2} m^2} \partial_\alpha \partial_{\beta'} I_\mu(\eta,\eta',\vec{x}-\vec{x}') \eqend{,} \label{space-space}
\end{align}
\end{subequations}
where we defined
\begin{equation}
\label{i1_def}
I_\nu(\eta,\eta',\vec{r}) = 2^{n-2} \pi^\frac{n+2}{2} (\eta\eta')^\frac{n-1}{2} \int \hankel1\nu{-\abs{\vec{p}} \eta} \hankel2\nu{-\abs{\vec{p}}\eta'} \mathe^{\mathi \vec{p} \vec{r}} \frac{\total^{n-1} p}{(2\pi)^{n-1}} \eqend{.}
\end{equation}
We have used the equality
\begin{equation}
\mathe^{\mathi\nu\frac{\pi}{2}} \hankel1\nu{x} \left[ \mathe^{\mathi \nu\frac{\pi}{2}} \hankel1\nu{x'} \right]^* = \hankel1\nu{x} \hankel2\nu{x'}\eqend{,}
\end{equation}
satisfied if $\nu$ is either real or purely imaginary and if $x$ and $x'$ are real. The integral \eqref{i1_def} is calculated in~\ref{appA} and depends only on the de~Sitter invariant $Z(x,x') = \cos(H\sigma(x,x'))$, where $\sigma(x,x')$ is the proper distance along the shortest geodesic connecting $x$ and $x'$ if these points are spacelike separated. The result is (see~\eqref{appendix_propagatorintegrals_i1_result})
\begin{equation}
\label{i1_z}
\begin{split}
I_\nu(Z) &= \frac{\Gamma\left( \frac{n-1}{2}+\nu \right) \Gamma\left( \frac{n-1}{2}-\nu \right)}{\Gamma\left( \frac{n}{2} \right)} \hypergeom{2}{1}\bigg( \frac{n-1}{2}+\nu, \frac{n-1}{2}-\nu; \frac{n}{2}; \frac{1+Z}{2} - \mathi \epsilon \sgn(\eta-\eta') \bigg) \eqend{.}
\end{split}
\end{equation}

In the conformally flat coordinate system of the Poincaré patch that we use, $Z$ is given by
\begin{equation}
Z(x,x') = 1 - \frac{(\vec{x}-\vec{x}')^2 - (\eta-\eta')^2}{2 \eta \eta'} \eqend{.}
\end{equation}
For its derivatives one easily finds the useful formulas
\begin{equation}
\label{z_time_der}
\partial_\eta Z = \frac{1}{\eta'} - \frac{Z}{\eta} \eqend{,} \qquad \partial_{\eta'} Z = \frac{1}{\eta} - \frac{Z}{\eta'} \eqend{,} \qquad \partial_\eta \partial_{\eta'} Z = - \frac{1}{\eta^2} - \frac{1}{(\eta')^2} + \frac{Z}{\eta\eta'}
\end{equation}
and
\begin{equation}
\label{z_spatial_der}
\partial_\alpha Z = - \frac{(\vec{x}-\vec{x}')_\alpha}{\eta\eta'} \eqend{,} \qquad \partial_{\beta'} Z = \frac{(\vec{x}-\vec{x}')_\beta}{\eta\eta'} \eqend{,} \qquad \partial_\alpha \partial_{\beta'} Z = \frac{\eta_{\alpha\beta}}{\eta\eta'} \eqend{.}
\end{equation}
For components of the two-point function with at least one temporal index $0$, we may use these expressions to evaluate the derivatives of $I_\nu(\eta,\eta',\vec{x}-\vec{x}')$
and obtain
\begin{subequations}
\label{temporal-components}
\begin{align}
\bra{0} A_0(x) A_0(x') \ket{0} &= \frac{H^{n-2}}{(4\pi)^\frac{n}{2} m^2} \bigg( \left( (n-1) I'_\nu(Z) + Z I''_\nu(Z) \right) \left( \partial_\eta Z \right) \left( \partial_{\eta'} Z \right) \nonumber \\
&\qquad+ \left( - (n-1) Z I'_\nu(Z) + (1-Z^2) I''_\nu(Z) \right) \partial_\eta \partial_{\eta'} Z - \partial_\eta \partial_{\eta'} I_\mu(Z) \bigg) \eqend{,} \\
\bra{0} A_0(x) A_{\beta'}(x') \ket{0} &= \frac{H^{n-2}}{(4\pi)^\frac{n}{2} m^2} \bigg( \left( (n-1) I'_\nu(Z) + Z I''_\nu(Z) \right) \left( \partial_\eta Z \right) \left( \partial_{\beta'} Z \right) \nonumber \\
&\qquad+ \left( - (n-1) Z I'_\nu(Z) + (1-Z^2) I''_\nu(Z) \right) \partial_\eta \partial_{\beta'} Z - \partial_\eta \partial_{\beta'} I_\mu(Z) \bigg) \eqend{,} \\
\bra{0} A_\alpha(x) A_0(x') \ket{0} &= \frac{H^{n-2}}{(4\pi)^\frac{n}{2} m^2} \bigg( \left( (n-1) I'_\nu(Z) + Z I''_\nu(Z) \right) \left( \partial_\alpha Z \right) \left( \partial_{\eta'} Z \right) \nonumber \\
&\qquad+ \left( - (n-1) Z I'_\nu(Z) + (1-Z^2) I''_\nu(Z) \right) \partial_\alpha \partial_{\eta'} Z - \partial_\alpha \partial_{\eta'} I_\mu(Z) \bigg) \eqend{.}
\end{align}
\end{subequations}
The evaluation of the space-space component \eqref{space-space} is more involved and will be delegated to Appendix~\ref{appB}, where we show that a result of the same form as \eqref{temporal-components} is obtained. Thus, canonical quantization of the Stueckelberg Lagrangian in the Bunch-Davies vacuum leads to the de~Sitter-invariant result
\begin{equation}
\label{wightman_result}
\bra{0} A_a(x) A_{b'}(x') \ket{0} = \frac{H^{n-2}}{(4\pi)^\frac{n}{2} m^2} \left( K_{ab'}(x,x') - \partial_a \partial_{b'} I_\mu(Z) \right)
\eqend{,}
\end{equation}
where
\begin{equation}
\begin{split}
K_{ab'}(x,x') &= \left( (n-1) I'_\nu(Z) + Z I''_\nu(Z) \right) \left( \partial_a Z \right) \left( \partial_{b'} Z \right) \\
&+ \left( - (n-1) Z I'_\nu(Z) + (1-Z^2) I''_\nu(Z) \right) \partial_a \partial_{b'} Z
\end{split}
\end{equation}
with $I_\nu(Z)$ defined by \eqref{i1_z}. The tensor $K_{ab'}(x,x')$ is transverse, which can readily be verified by using the relations
\begin{equation}
\nabla^a \nabla_a Z = - n H^2 Z \eqend{,} \qquad \left( \nabla^a Z \right) \nabla_a Z = H^2 (1-Z^2) \eqend{,} \qquad \left( \nabla^a Z \right) \nabla_a \nabla_{b'} Z = - H^2 Z \nabla_{b'} Z \eqend{,}
\end{equation}
while the term $-\partial_a \partial_{b'} I_\mu(Z)$ can be identified as the longitudinal part. Note that
\begin{equation}
\label{M1}
\Delta_{\xi m^2}(Z) = \frac{H^{n-2}}{(4\pi)^{\frac{n}{2}}} I_\mu(Z)
\end{equation}
is the two-point function of a minimally coupled scalar field with squared mass $M^2=\xi m^2$\ \cite{candelasraine1975}. Eq.~\eqref{wightman_result} is valid for all $n\geq 2$, since the contribution in~\eqref{AA} from $B_\alpha$ vanishes automatically if $n=2$.

In odd dimensions the function $I_\nu(Z)$ can be expressed in terms of elementary functions as follows. First we recall
\begin{equation}
\label{easy}
\hypergeom{2}{1}\left( \nu, -\nu; \frac{1}{2}; \frac{1-\cos y}{2} \right) = \cos (\nu y)
\end{equation}
if $|y|<\pi$. This formula can readily be verified by noting that the hypergeometric equation satisfied by the left-hand side reduces to the simple second-order differential equation satisfied by the right-hand side, and that both satisfy the same initial conditions at $y=0$.
From the explicit expression of $I_\nu(Z)$~\eqref{i1_z} it is seen that by differentiating it with respect to $Z$ we obtain $I_\nu(Z)$ in a higher dimension:
\begin{equation}
\label{Z-derivative}
\left. 2 \frac{\total}{\total Z} I_\nu(Z) \right|_n = \left. I_\nu(Z) \right|_{n+2} \eqend{.}
\end{equation}
For $n=3$ the use of~\eqref{easy} and \eqref{Z-derivative} with $Z=-\cos y$ gives
\begin{equation}
\label{I_nu_3D}
\left. I_\nu(Z) \right|_{n=3} = \frac{2\sqrt{\pi}}{\sin(\pi \nu)} \frac{\sin (\nu y)}{\sin y} \eqend{,}
\end{equation}
and for general odd dimensions $n=2k+1$, $k=1,2,\ldots$, we find
\begin{equation}
\label{I_nu_oddD}
\left. I_\nu(Z) \right|_{n=2k+1} = - \frac{\sqrt{\pi}}{\nu \sin(\pi \nu)} \left( \frac{2}{\sin y} \frac{\total}{\total y} \right)^k \cos (\nu y) \eqend{.}
\end{equation}
If the two points $x$ and $x'$ are spacelike separated, then $y = \pi - H\sigma$, where $\sigma$ is the geodesic distance between $x$ and $x'$.

\section{Limits of zero mass and zero and infinite Stueckelberg parameter}
\label{limits}

The zero-mass limit of the Proca theory (with $\xi \to \infty$ in \eqref{st_lag}) is singular because the condition $\nabla^a A_a = 0$ arises only if $m \neq 0$. On the other hand, in the Stueckelberg theory with finite value of $\xi$ the zero-mass limit exists and is the standard massless vector theory with the covariant gauge-fixing term. We verify this fact explicitly by taking the zero-mass limit of the Wightman two-point function \eqref{wightman_result}.

It is convenient to write \eqref{wightman_result} as
\begin{equation}
\label{AA-first}
\bra{0} A_a(x) A_{b'}(x') \ket{0} = \frac{H^{n-4}}{(4\pi)^\frac{n}{2}} \Big[ A(Z) \partial_a \partial_{b'} Z + B(Z) \left( \partial_a Z \right) \left( \partial_{b'} Z \right) \Big] \eqend{,}
\end{equation}
where
\begin{subequations}
\label{AB}
\begin{align}
A(Z) &= \frac{H^2}{m^2} \left( F_\nu(Z) - I'_\mu(Z) \right) + I_\nu(Z) \eqend{,} \\
B(Z) &= \frac{H^2}{m^2} \left( F'_\nu(Z) - I''_\mu(Z) \right) \eqend{,}
\end{align}
\end{subequations}
with
\begin{equation}
F_\nu(Z) = Z I_\nu'(Z) + (n-2) I_\nu(Z) \eqend{.}
\end{equation}
In deriving this form, we have used the hypergeometric equation satisfied by $I_\nu(Z)$ as a consequence of the equations of motion:
\begin{equation}
\label{simple-hyper}
(1-Z^2) I_\nu''(Z) = n Z I_\nu'(Z) + \left( n-2 + \frac{m^2}{H^2} \right) I_\nu(Z) \eqend{.}
\end{equation}
By using identities satisfied among hypergeometric functions with different indices, or by term-by-term comparison of the series expressions, we find
\begin{equation}
F_\nu(Z) = G_\nu'(Z) + \left( \frac{n-3}{2} - \nu \right) I_\nu(Z) \eqend{,}
\end{equation}
where
\begin{equation}
\label{g_nu_def}
G_\nu(Z) = \frac{\Gamma\left( \frac{n+1}{2} + \nu \right) \Gamma\left( \frac{n-3}{2} - \nu \right)}{\Gamma\left( \frac{n}{2} \right)} \hypergeom{2}{1}\left( \frac{n+1}{2} + \nu, \frac{n-3}{2} - \nu; \frac{n}{2}; \frac{1+Z}{2} - \mathi \epsilon \sgn(\eta-\eta') \right) \eqend{.}
\end{equation}
Now we assume that $n \geq 4$. Then
\begin{equation}
\lim_{m^2\to 0} \frac{H^2}{m^2} \left( \frac{n-3}{2} - \nu \right) = \frac{1}{n-3} \eqend{.}
\end{equation}
Thus, we find from \eqref{AA-first}
\begin{equation}
\label{AA-massless}
\begin{split}
\lim_{m^2\to 0} \bra{0} A_a(x) A_{b'}(x') \ket{0} &= \frac{H^{n-4}}{(4\pi)^\frac{n}{2}} \left[ \frac{n-2}{n-3} I^{(0)}(Z) \partial_a \partial_{b'} Z + \frac{1}{n-3} I^{(0)\prime}(Z) \left( \partial_a Z \right) \left( \partial_{b'} Z \right) \right] \\
&\quad+ \frac{H^{n-4}}{(4\pi)^\frac{n}{2}} \lim_{m^2 \to 0} \frac{H^2}{m^2} \partial_a \partial_{b'} \left( G_\nu(Z) - I_\mu(Z) \right) \eqend{,}
\end{split}
\end{equation}
where
\begin{equation}
I^{(0)}(Z) = \lim_{m^2 \to 0} I_\nu(Z) = \frac{\Gamma(n-2)}{\Gamma\left( \frac{n}{2} \right)} \hypergeom{2}{1}\left( n-2, 1; \frac{n}{2}; \frac{1+Z}{2} - \mathi\epsilon\sgn(\eta-\eta') \right) \eqend{.}
\end{equation}
If we define the mass $M$ by
\begin{equation}
\frac{M^2}{H^2} = \left( \frac{n-1}{2} \right)^2 - (\nu + 1)^2 = \frac{n-1}{n-3} \frac{m^2}{H^2} + \bigo{m^4} \eqend{,}
\end{equation}
the scalar propagator $\Delta_{M^2}$ is proportional to $G_\nu(Z)$~\eqref{g_nu_def}
\begin{equation}
\label{M2}
\Delta_{M^2}(Z) = \frac{H^{n-2}}{(4\pi)^\frac{n}{2}} G_\nu(Z) \eqend{.}
\end{equation}
In the small mass limit $M^2 \to 0$, this propagator is divergent, but the divergent contribution is $Z$-independent and drops out when the derivatives are taken in~\eqref{AA-massless}. The $m$-independent term cancels between $G_\nu(Z)$ and $I_\mu(Z)$ (we recall~\eqref{M1} that $I_\mu(Z)$ is proportional to $\Delta_{\xi m^2}$), and we extract the terms linear in $m^2$ by defining
\begin{equation}
\tilde{\Delta}(Z) = - \lim_{m^2 \to 0} \frac{\partial}{\partial m^2} \left( \Delta_{m^2}(Z) - \Delta_{m^2}(-1) \right) \eqend{.}
\end{equation}
The massless limit of our two-point function is then given by
\begin{equation}
\label{m-less-limit}
\begin{split}
\lim_{m^2\to 0} \bra{0} A_a(x) A_{b'}(x') \ket{0} &= \frac{H^{n-4}}{(4\pi)^\frac{n}{2}} \left[ \frac{n-2}{n-3} I^{(0)}(Z) \partial_a \partial_{b'} Z + \frac{1}{n-3} I^{(0)\prime}(Z) \left( \partial_a Z \right) \left( \partial_{b'} Z \right) \right] \\
&\quad+ \left( \xi - \frac{n-1}{n-3} \right) \partial_a \partial_{b'} \tilde{\Delta}(Z) \qquad (n\geq 4) \eqend{.}
\end{split}
\end{equation}
For $n = 4$ this result agrees with that found by Youssef~\cite{youssef2010} extending the $\xi=1$ case discussed by Allen and Jacobson~\cite{allenjacobson1986}, as shown in Appendix~\ref{appC}. In this Appendix we also present the massless limit for $n = 2$ and $n = 3$.

The limit $\xi \to 0$ (known as Landau gauge) is straightforward. (Note that we let $m^2 > 0$ again.) We only need to take the limit $\mu \to \frac{n-1}{2}$, hence, the two-point function is given by~\eqref{wightman_result} with the following substitution:
\begin{equation}
\label{tsamwood}
\partial_a \partial_{b'} I_\mu(Z) \to I^{(1)}(Z) \partial_a \partial_{b'} Z + I^{(1)\prime}(Z) \left( \partial_a Z \right) \left( \partial_{b'} Z \right) \eqend{,}
\end{equation}
where
\begin{equation}
\label{zeromass_imu}
I^{(1)}(Z) = \lim_{\mu \to \frac{n-1}{2}} I'_\mu(Z) = \frac{\Gamma(n)}{2\Gamma\left( \frac{n}{2}+1 \right)} \hypergeom{2}{1}\left( n, 1; \frac{n}{2}+1; \frac{1+Z}{2} - \mathi \epsilon \sgn(\eta-\eta') \right) \eqend{.}
\end{equation}

The limit $\xi \to \infty$  (known as the unitary gauge) is not as straightforward. Here we only study the case where the two points $x$ and $x'$ can be connected by a spacelike geodesic, i.e.\ where $0 < \sigma < \pi$. If $n$ is odd, then $I_\mu(Z)$ is given by \eqref{I_nu_oddD}. For large $\xi$ we have $\mu = \mathi \kappa$, where
\begin{equation}
\kappa = \sqrt{\xi \frac{m^2}{H^2} - \left( \frac{n-3}{2} \right)^2} \eqend{.}
\end{equation}
(Notice that $I_{\mathi\kappa}(Z) = I_{-\mathi\kappa}(Z)$, so the sign of $\kappa$ does not matter.) Then
\begin{equation}
\frac{\sin (\nu y)}{\sin (\pi \nu)}  =  \frac{\sinh[ \kappa (\pi - \sigma) ]}{\sinh (\pi \kappa)} \approx \mathe^{-\kappa \sigma}  \eqend{.}
\end{equation}
This implies that $I_\mu(Z)$ tends to zero exponentially as $\xi \to \infty$ if $n$ is odd. The same conclusion can be drawn for $n$ even by using the series expression of $I_\mu(Z)$:
\begin{equation}
I_\mu(Z) = \sum_{k=0}^\infty \frac{\abs{\Gamma\left( \frac{n-1}{2}+k+\mathi\kappa \right)}^2}{\Gamma\left( \frac{n}{2}+k \right) k!} z^k \eqend{,}
\end{equation}
where $z = (1+Z)/2$ and $0 < z < 1$. If $n \geq 2$ is even, then by defining $n_+ = n+1$ we find
\begin{subequations}
\label{pi-pi}
\begin{align}
\abs{\Gamma\left( \frac{n-1}{2}+k+\mathi\kappa \right)}^2 &\leq \pi \abs{\Gamma\left( \frac{n_+-1}{2}+k+\mathi\kappa \right)}^2 \eqend{,} \\
\Gamma\left( \frac{n}{2}+k \right) &\geq \pi^{-\frac{1}{2}} \Gamma\left( \frac{n_+}{2}+k-1 \right) \eqend{.}
\end{align}
\end{subequations}
The second inequality holds without the factor $\pi^{-\frac{1}{2}}$ if $(n,k)\neq (2,0)$.  Hence
\begin{equation}
\abs{I_\mu(Z)} \leq \pi^\frac{3}{2} \sum_{k=0}^\infty \frac{\abs{\Gamma\left( \frac{n_+-1}{2}+k+\mathi\kappa \right)}^2}{\Gamma\left( \frac{n_+}{2}+k \right) k!} \left( \frac{n_+}{2}+k \right) z^k = \pi^\frac{3}{2} \abs{ z^{-\frac{n_+}{2}} \frac{\total}{\total z} \left( z^\frac{n_+}{2} \left. I_\mu(Z) \right|_{n\to n_+} \right) } \eqend{,}
\end{equation}
where $\left. I_\mu(Z) \right|_{n\to n_+}$ is the function obtained from $I_\mu(Z)$ by keeping $\mu = \mathi\kappa$ unchanged but replacing $n$ by $n_+=n+1$ everywhere else. Since $\left. I_\mu(Z) \right|_{n\to n_+}$ tends to zero exponentially as $\xi\to 0$, we conclude that $I_\mu(Z)$ does so as well.

Thus, in the limit $\xi\to \infty$ the two-point function is purely transverse, as it must be for the Proca theory, where $\nabla^a A_a = 0$ follows from the equation of motion. However, now the limit $m \to 0$ diverges like $\sim 1/m^2$ since $I_\mu(Z)$ vanishes and cannot cancel the divergence arising from the other terms in \eqref{wightman_result} (at least for spacelike separated points). This is the \emph{expected} behavior that also appears in flat space, as can be seen in the next section.

\section{The flat-space limit}
\label{flatspacelimit}

The flat-space two-point function is recovered in the limit of infinite de~Sitter radius $H \to 0$. For the flat-space limit we may set $x'=0$ because of Lorentz invariance. Then to lowest order in $H$ we have
\begin{equation}
Z \approx 1 - \frac{1}{2} H^2 x^2 \eqend{,} \qquad \frac{\partial}{\partial Z} \approx - \frac{2}{H^2} \frac{\partial}{\partial x^2} \eqend{,} \qquad \nu \approx \mathi \frac{m}{H} \eqend{.}
\end{equation}
Using the approximation $\eta \approx - 1/H + t$, the analytic continuation $Z \to Z - \mathi \epsilon \sgn(\eta-\eta')$ results in $x^2 \to x^2 + \mathi \epsilon \sgn t$. We will now suppress this continuation and let it be understood implicitly.

To obtain the flat-space limit of $I_\nu(Z)$ from Eq.~\eqref{i1_z}, we first have to use a hypergeometric transformation sending $z=(1+Z)/2$ to $1-z$ to obtain
\begin{equation}
\label{i1_z_1mz}
\begin{split}
I_\nu(Z) &= \frac{\pi}{\sin\left( \frac{n}{2} \pi \right)} \frac{\Gamma\left( \frac{n-1}{2} + \nu \right) \Gamma\left( \frac{n-1}{2} - \nu \right)}{\Gamma\left( \frac{1}{2} + \nu \right) \Gamma\left( \frac{1}{2} - \nu \right) \Gamma\left( \frac{n}{2} \right)} \hypergeom{2}{1}\left( \frac{n-1}{2} + \nu, \frac{n-1}{2} - \nu, \frac{n}{2}, \frac{1-Z}{2} \right) \\
&\qquad+ \Gamma\left( \frac{n-2}{2} \right) \left( \frac{1-Z}{2} \right)^\frac{2-n}{2} \hypergeom{2}{1}\left( \frac{1}{2} + \nu, \frac{1}{2} - \nu; \frac{4-n}{2}, \frac{1-Z}{2} \right) \eqend{.}
\end{split}
\end{equation}
(For $n$ even we need to let $n$ have a non-integer value and let $n$ tend to the even integer in the end.)
We can then insert the series definition for the hypergeometric functions and take the flat-space limit term by term, using that
\begin{equation}
\frac{\Gamma(a+\nu) \Gamma(a-\nu)}{\Gamma(b+\nu) \Gamma(b-\nu)} \approx \left( \frac{m^2}{H^2} \right)^{a-b} \eqend{.}
\end{equation}
This gives
\begin{equation}
\begin{split}
I_\nu(Z) \approx \frac{\pi}{\sin\left( \frac{n}{2} \pi \right)} \left( \frac{m^2}{H^2} \right)^{\frac{n-2}{2}} \Bigg[ &\frac{1}{\Gamma\left( \frac{n}{2} \right)} \hypergeom{0}{1}\left( -; \frac{n}{2}; \frac{1}{4} m^2 x^2 \right) \\
&\quad- \frac{1}{\Gamma\left( \frac{4-n}{2} \right)} \left( \frac{1}{4} m^2 x^2 \right)^\frac{2-n}{2} \hypergeom{0}{1}\left( -; \frac{4-n}{2}; \frac{1}{4} m^2 x^2 \right) \Bigg] \eqend{.}
\end{split}
\end{equation}
Recognizing this series as a sum of two modified Bessel functions, we find
\begin{equation}
I_\nu(Z) \approx 2 \left( \frac{H^2}{2m^2} \right)^\frac{2-n}{2} \left( m \sqrt{x^2} \right)^\frac{2-n}{2} \besselk{\frac{n-2}{2}}{m \sqrt{x^2}} \eqend{.}
\end{equation}
The same calculation for $I_\mu(Z)$ yields
\begin{equation}
\label{I-mu-flat}
I_\mu(Z) \approx 2 \left( \frac{H^2}{2\xi m^2} \right)^\frac{2-n}{2} \left( m \sqrt{\xi x^2} \right)^\frac{2-n}{2} \besselk{\frac{n-2}{2}}{m \sqrt{\xi x^2}} \eqend{.}
\end{equation}
For the two-point function~\eqref{wightman_result} we therefore obtain
\begin{equation}
\label{flatspace_wightman}
\begin{split}
\lim_{H\to 0}\bra{0} A_a(x) A_{b'}(x') \ket{0} &= \frac{m^{n-4}}{(2\pi)^\frac{n}{2}} \Bigg[ - 4 x_a x_{b'} \frac{\partial^2}{\partial (x^2)^2} F_n\left( m \sqrt{x^2} \right) + \xi^\frac{n-2}{2} \partial_a \partial_{b'} F_n\left( m \sqrt{\xi x^2} \right) \\
&\qquad+ 2 \eta_{ab'} \left( (n-1) \frac{\partial}{\partial x^2} F_n\left( m \sqrt{x^2} \right) + 2 x^2 \frac{\partial^2}{\partial (x^2)^2} F_n\left( m \sqrt{x^2} \right) \right) \Bigg] \\
&= \frac{m^{n-4}}{(2\pi)^\frac{n}{2}} \Bigg[ \left( \xi^\frac{n-2}{2} - 1 \right) \partial_a \partial_{b'} F_n\left( m \sqrt{\xi x^2} \right) + \eta_{ab'} m^2 F_n\left( m \sqrt{x^2} \right) \Bigg]
\end{split}
\end{equation}
with
\begin{equation}
\label{def-F}
F_n(z) = z^\frac{2-n}{2} \besselk{\frac{n-2}{2}}{z} \eqend{.}
\end{equation}
Here we are denoting
$\partial/\partial x^{b'}$, rather than $\partial/\partial (x')^{b'} = - \partial/\partial x^{b'}$, by $\partial_{b'}$ unlike elsewhere in this paper.
Recall that we have set $x' = 0$, so that (in this section)
\begin{equation}
\partial_a \partial_{b'} f(x^2) = 2 \eta_{ab'} \frac{\partial f(x^2)}{\partial x^2} + 4 x_a x_{b'} \frac{\partial^2 f(x^2)}{\partial(x^2)^2}
\end{equation}
for any function $f$. We also have used the following equation satisfied by $F_n$ (which is essentially the Klein-Gordon equation in Minkowski space):
\begin{equation}
\partial_a \partial^a F_n(m\sqrt{x^2}) = m^2 F_n(m\sqrt{x^2}) \eqend{.}
\end{equation}
We already see that for Feynman gauge $\xi = 1$ we are left with a simpler result proportional to $\eta_{ab'}$ for the flat-space limit of the two-point function.

As we stated before, the continuation $x^2\to x^2 + \mathi \epsilon \sgn t$ was implicitly understood above for the Wightman function. Next, we are going to verify that the Feynman propagator with $x^2\to x^2 + \mathi \epsilon$ (hence $Z\to Z - \mathi\epsilon$) obtained from \eqref{flatspace_wightman} agrees with the well-known result. We note that the Feynman propagator for the minimally-coupled scalar field, $\Delta_{m^2}(Z-\mathi\epsilon)$~\eqref{M1}, should tend to the well-known Feynman propagator in the flat-space limit, i.e.
\begin{equation}
\lim_{H \to 0} \Delta_{m^2}(Z - \mathi \epsilon) = - \mathi \int \frac{\mathe^{\mathi p x}}{p^2 + m^2 - \mathi \epsilon} \frac{\total^n p}{(2\pi)^n} \eqend{.}
\end{equation}
This equation and \eqref{I-mu-flat} imply
\begin{equation}
\label{F-check}
F_n\left( m \sqrt{x^2 + \mathi \epsilon} \right) = - \mathi (2\pi)^\frac{n}{2} m^{2-n} \int \frac{\mathe^{\mathi p x}}{p^2 + m^2 - \mathi \epsilon} \frac{\total^n p}{(2\pi)^n} \eqend{,}
\end{equation}
where $F_n(z)$ is defined by \eqref{def-F}. Now, we verify this expression explicitly. We need the Fourier transform for a complex power of the massless propagator (Eq.~(A.40) in Ref.~\onlinecite{smirnovfeynman}, converted to our conventions)
\begin{equation}
\frac{1}{(x^2 + \mathi \epsilon)^s} = - \mathi \frac{(4\pi)^\frac{n}{2} \Gamma\left( \frac{n}{2}-s \right)}{4^s \Gamma(s)} \int \frac{\mathe^{\mathi p x}}{(p^2 - \mathi \epsilon)^{\frac{n}{2}-s}} \frac{\total^n p}{(2\pi)^n} \eqend{,}
\end{equation}
where $0 < s < \frac{n}{2}$, and the Mellin integral representation of the modified Bessel function
\begin{equation}
\besselk{\alpha}{z} = \frac{1}{2} \int_{c-\mathi\infty}^{c+\mathi\infty} \Gamma(s) \Gamma(s-\alpha) \left( \frac{z}{2} \right)^{\alpha-2s} \frac{\total s}{2\pi\mathi}
\end{equation}
with $c > \Re(\alpha), c > 0$, where $\Re(\alpha)$ is the real part of $\alpha$. Notice the sign difference of the $\mathi \epsilon$ term between coordinate and momentum spaces. One then readily obtains
\begin{equation}
F_n\left( m \sqrt{x^2+\mathi\epsilon} \right) = \mathi (2\pi)^\frac{n}{2} m^{-n} \int_{c-\mathi\infty}^{c+\mathi\infty} \frac{\pi}{\sin\left[ \pi\left( s-\frac{n}{2} \right) \right]} \int \left( \frac{m^2}{p^2 - \mathi \epsilon} \right)^{\frac{n}{2}-s} \mathe^{\mathi p x} \frac{\total^n p}{(2\pi)^n} \frac{\total s}{2\pi\mathi}
\end{equation}
with $\frac{n-2}{2} < c < \frac{n}{2}$. Interchanging the order of integration, we can close the $s$ integration contour to the right if $\abs{p^2} < m^2$ and to the left if $\abs{p^2} > m^2$, and sum the residues to obtain~\eqref{F-check}.
The Feynman propagator $G^\mathrm{F}_{ab'}(x-x') = - \mathi \bra{0} \mathcal{T} A_a(x) A_{b'}(x') \ket{0}$, where the time-ordering $\mathcal{T}$ has a correction term (see the next section), is then obtained from the two-point function~\eqref{flatspace_wightman} with the Feynman prescription and given by
\begin{equation}
\label{feynman_flatspace}
\begin{split}
G^\mathrm{F}_{ab'}(x) &= - \int \Bigg[ \left( \eta_{ab'} + \frac{p_a p_{b'}}{m^2} \right) \frac{\mathe^{\mathi p x}}{p^2 + m^2 - \mathi \epsilon} - \frac{p_a p_{b'}}{m^2} \frac{\mathe^{\mathi p x}}{p^2 + \xi m^2 - \mathi \epsilon} \Bigg] \frac{\total^n p}{(2\pi)^n} \\
&= - \int \Bigg[ \eta_{ab'} \frac{\mathe^{\mathi p x}}{p^2 + m^2 - \mathi \epsilon} + p_a p_{b'} (\xi-1) \frac{\mathe^{\mathi p x}}{\left( p^2 + m^2 - \mathi \epsilon \right) \left( p^2 + \xi m^2 - \mathi \epsilon \right)} \Bigg] \frac{\total^n p}{(2\pi)^n} \eqend{.} \\
\end{split}
\end{equation}
The second form is suitable for taking the limit $m \to 0$. This is the well-known expression for the Feynman propagator in flat space~\cite{itzyksonzuber}, including the prescription of adding a negative imaginary part to the squared mass: $m^2 \to m^2 - \mathi \epsilon$. It can also be seen very clearly that both limits $m \to 0$ and $\xi \to \infty$ can not be taken: after taking the limit $\xi \to \infty$ which makes the second term in the first line vanish, the propagator diverges like $\sim 1/m^2$, while after taking the limit $m \to 0$ in the second line the propagator diverges linearly as $\xi \to \infty$.

\section{The divergence of the Feynman propagator}
\label{divergencefeynman}

In this section we discuss the divergence of the Feynman propagator in the Proca and Stueckelberg theories.  We note that Tsamis and Woodard claimed
erroneously~\cite{tsamiswoodard2007} that the propagator of Allen and Jacobson~\cite{allenjacobson1986} was wrong because its divergence is nonzero.
Most of the facts explained here are well known in the flat-space limit.

For the strict Proca theory (i.e., in the theory obtained by taking the limit $\xi \to \infty$ in the Lagrangian~\eqref{st_lag} before quantization), the zero component of the vector field $A_0$ is not an independent degree of freedom. Rather, the equations of motion determine it as
\begin{equation}
\label{proca_a0_eom}
A_0 = - \frac{(-H\eta)^{n-2}}{m^2} \partial_\alpha \pi^\alpha \eqend{.}
\end{equation}
This implies that the equal-time commutator $[ A_\alpha(\vec{x},\eta), A_0(\vec{x}',\eta) ]$ does not vanish. This fact leads to the non-vanishing of the divergence of the Feynman propagator as we shall see below. This commutator is analogous to the non-vanishing commutator of the space and time components of conserved current, known as
the ``Schwinger term''~\cite{schwinger1959,boulwaredeser1967,jacobstech1970}. We also note that the naive definition of time ordering leads to a non-covariant Feynman propagator. For covariance it must be supplemented by a local term~\cite{dashenlee1969,nutbrown1971,grossjackiw1969} so that we have
\begin{equation}
\begin{split}
G^\mathrm{F}_{ab'}(x,x') &= - \mathi \left( \Theta(\eta-\eta') \bra{0} A_a(x) A_{b'}(x') \ket{0} + \Theta(\eta'-\eta) \bra{0} A_{b'}(x') A_a(x) \ket{0} \right) \\
&\quad+ \frac{1}{m^2} \delta^0_a \delta^0_{b'} (-H\eta)^{n-2} \delta^n(x-x') \eqend{.}
\end{split}
\end{equation}
Noting that the vector field is transverse in the Proca theory $\nabla^a A_a = 0$, we readily obtain
\begin{equation}
\nabla^a G^\mathrm{F}_{ab'}(x,x') = \mathi (-H\eta)^2 \delta(\eta-\eta') \bra{0} [ A_0(x), A_{b'}(x') ] \ket{0} - \frac{1}{m^2} \delta^0_{b'} (-H\eta)^n \partial_\eta \delta^n(x-x') \eqend{,}
\end{equation}
and the equal-time commutator does not vanish. Rather, using~\eqref{proca_a0_eom} and the canonical commutation relations, we find
\begin{equation}
\nabla^a G^\mathrm{F}_{ab'}(x,x') = \delta(\eta-\eta') \frac{(-H\eta)^n}{m^2} \partial_{b'} \delta^{n-1}(\vec{x}-\vec{x}') - \frac{1}{m^2} \delta^0_{b'} (-H\eta)^n \partial_\eta \delta^n(x-x') \eqend{,}
\end{equation}
which can be combined into the covariant form
\begin{equation}
\label{divergenceG}
\nabla^a G^\mathrm{F}_{ab'}(x,x') = \frac{1}{m^2} \partial_{b'} \left( \frac{\delta^n(x-x')}{\sqrt{-g}} \right) \eqend{.}
\end{equation}
Note that the additional local term added to the definition of the time-ordered two-point function indeed serves to make the divergence of the Feynman propagator covariant~\cite{dashenlee1969,nutbrown1971,grossjackiw1969}. It is known that this term does not affect observables~\cite{grossjackiw1969,brown1967}.

In contrast, for the Stueckelberg Lagrangian the field $A_0$ represents an independent degree of freedom, and the equal-time commutator $[ A_0(\vec{x},\eta), A_\alpha(\vec{x}',\eta) ]$ vanishes. In this case, one also finds that the simple definition of time ordering is sufficient to obtain a covariant Feynman propagator. As we have noted above, there is a local term in the divergence of the Feynman propagator for the transverse part. Let $\phi_{M^2}(x)$ be the field operator for the minimally-coupled scalar field with mass $M$ with the standard commutation relations and let $\ket{0}$ be the Bunch-Davies vacuum for this field. Then the Feynman propagator of the longitudinal part can be written as
\begin{equation}
\begin{split}
G^\mathrm{F(L)}_{ab'}(x,x') &= \frac{\mathi}{m^2} \left[
\Theta(\eta-\eta') \bra{0} \left( \partial_a \phi_{\xi m^2}(x) \right) \left( \partial_{b'} \phi_{\xi m^2}(x') \right) \ket{0} \right. \\
&\qquad\quad\left. + \Theta(\eta'-\eta) \bra{0} \left( \partial_{b'} \phi_{\xi m^2}(x') \right) \left( \partial_a \phi_{\xi m^2}(x) \right) \ket{0} \right] \eqend{.}
\end{split}
\end{equation}
The divergence of this propagator is
\begin{equation}
\begin{split}
\nabla^a G^\mathrm{F(L)}_{ab'}(x,x') &= - \delta(\eta-\eta') \frac{(-H\eta)^n}{m^2} \partial_{b'} \delta^{n-1}(\vec{x}-\vec{x}') \\
&\qquad+ \mathi \xi \frac{H^{n-2}}{(4\pi)^\frac{n}{2}} \left( \Theta(\eta-\eta') \nabla_{b'} I_\mu(Z) + \Theta(\eta'-\eta) \nabla_{b'} I^*_\mu(Z) \right) \eqend{.}
\end{split}
\end{equation}
In the full Feynman propagator the first term, which is local, cancels out the local term from the Feynman propagator of the transverse (Proca) part, and we are left with
\begin{equation}
\nabla^a G^\mathrm{F}_{ab'}(x,x') = \mathi \xi \nabla_{b'} \Delta_{\xi m^2}(Z-\mathi\epsilon) \eqend{,}
\end{equation}
which vanishes in the limit $\xi \to 0$ as expected.  The cancellation of the local terms occurs also if we let $Z\to Z-\mathi\epsilon$ for the Proca and longitudinal contributions.
Of course, this result should be consistent with the divergence of the Feynman propagator~\eqref{feynman_flatspace} in flat space. Indeed we obtain
\begin{equation}
\label{divG-eq}
\partial^a G^\mathrm{F}_{ab'}(x-x') = \xi \partial_{b'} \int \frac{\mathe^{\mathi p (x-x')}}{p^2 + \xi m^2 - \mathi \epsilon} \frac{\total^n p}{(2\pi)^n} \eqend{.}
\end{equation}
We note that this gives $1/m^2\, \partial_{b'} \delta^n(x-x')$ as $\xi \to \infty$, which agrees with the result in the Proca theory.

\section{Comparison with earlier results}
\label{comparison}

To compare our results with the ones by Allen and Jacobson~\cite{allenjacobson1986}, we first have to convert our results to their notation. As a basis of bi-tensors, Allen and Jacobson use the normal vectors to the spacelike geodesic connecting $x$ and $x'$ with geodesic length $\sigma(x,x')$ defined by
\begin{equation}
n_a = \nabla_a \sigma \eqend{,} \qquad n_{a'} = \nabla_{a'} \sigma
\end{equation}
and the parallel propagator $g_{ab'}$. Since $Z = \cos (H \sigma)$, we immediately find the relation
\begin{equation}
\nabla_a Z = - H \sin(H \sigma) n_a \eqend{,} \qquad \nabla_{a'} Z = - H \sin(H \sigma) n_{a'} \eqend{.}
\end{equation}
Furthermore, by means of the identity $\nabla_a n_{b'} = - H ( g_{ab'} + n_a n_{b'} )/\sin(H \sigma)$, one obtains
\begin{equation}
\nabla_a \nabla_{b'} Z = H^2 g_{ab'} + H^2 ( 1 - \cos(H \sigma) ) n_a n_{b'} \eqend{.}
\end{equation}
They use $z = \cos^2(H\sigma/2) = (1+Z)/2$ instead of $Z$. The two-point function of the vector field~\eqref{wightman_result} in the Allen-Jacobson notation reads
\begin{equation}
\begin{split}
\bra{0} A_a(x) A_{b'}(x') \ket{0} &= \frac{H^{n-2}}{(4\pi)^\frac{n}{2} m^2} \bigg[ - \partial_a \partial_{b'} I_\mu(z) \\
&\qquad- \frac{H}{2} n_a n_{b'} \left( H (n-1) (\cos(H\sigma)-1) + \sin(H\sigma) \partial_\sigma \right) \partial_z I_\nu(z) \\
&\qquad\left. - \frac{H}{2} g_{ab'} \left( H (n-1) \cos(H\sigma) + \sin(H\sigma) \partial_\sigma \right) \partial_z I_\nu(z) \right]
\end{split}
\end{equation}
with
\begin{equation}
I_\nu(z) = \frac{\Gamma\left( \frac{n-1}{2}+\nu \right) \Gamma\left( \frac{n-1}{2}-\nu \right)}{\Gamma\left( \frac{n}{2} \right)} \hypergeom{2}{1}\left( \frac{n-1}{2}+\nu, \frac{n-1}{2}-\nu; \frac{n}{2}; z - \mathi \epsilon \sgn(\eta-\eta') \right) \eqend{.}
\end{equation}
In the limit $\xi \to \infty$ (the unitary gauge) where $I_\mu(z) \to 0$ (at least for spacelike separated points), this coincides with the result by Allen and Jacobson except that they erroneously state that the Feynman propagator is obtained by letting $z \to z + \mathi \epsilon$ rather than $z \to z - \mathi \epsilon$. Since in this limit we recover the Proca theory, we agree with their Proca two-point function except for their sign error for the $\mathi \epsilon$ (see section~\ref{flatspacelimit}).

To compare our results with those by Tsamis and Woodard~\cite{tsamiswoodard2007}, we find it simpler to convert their notation into ours. Instead of $Z$ they use $y = 4 \sin^2(H\sigma/2) = 2 (1-Z)$ and the corresponding covariant derivatives. Their result can then be presented as follows
\begin{equation}
\begin{split}
\bra{0} A_a(x) A_{b'}(x') \ket{0} &= - \frac{1}{(n-1) H^2} \left( (1-Z^2) \gamma'(Z) - (n-1) Z \gamma(Z) \right) \partial_a \partial_{b'} Z \\
&\qquad- \frac{1}{(n-1) H^2} \left( Z \gamma'(Z) + (n-1) \gamma(Z) \right) \left( \partial_a Z \right) \left( \partial_{b'} Z \right) \eqend{,}
\end{split}
\end{equation}
where $\gamma(Z)$ is given by
\begin{equation}
\begin{split}
\gamma(Z) &= - \frac{n-1}{2} \frac{H^2}{m^2} \frac{H^{n-2}}{(4\pi)^\frac{n}{2}} \Bigg[ - \frac{\Gamma(n-1)}{\Gamma\left( \frac{n}{2}+1 \right)} \hypergeom{2}{1}\left( n-1, 2; \frac{n}{2}+1; \frac{1+Z}{2} \right) \\
&\qquad+ \frac{\Gamma\left( \frac{n+1}{2}+\nu \right) \Gamma\left( \frac{n+1}{2}-\nu \right)}{\Gamma\left( \frac{n}{2}+1 \right)} \hypergeom{2}{1}\left( \frac{n+1}{2}+\nu, \frac{n+1}{2}-\nu; \frac{n}{2}+1; \frac{1+Z}{2} \right) \Bigg] \eqend{.}
\end{split}
\end{equation}
While the second hypergeometric function corresponds to the pure Proca theory and yields the result by Allen and Jacobson, the first hypergeometric function gives a term proportional to $\partial_a \partial_{b'} I^{(1)}(Z)$, where the function $I^{(1)}(Z)$ is defined by~\eqref{zeromass_imu}, in the limit $\xi \to 0$ by expanding the hypergeometric functions around $Z = -1$. Therefore, while Tsamis and Woodard claim to have calculated the propagator for the pure Proca theory, in reality they obtained the Stueckelberg propagator in the limit of vanishing Stueckelberg parameter $\xi \to 0$. The reason for this unwarranted claim may be traced back to their wrong assertion that the Feynman propagator is transverse for the strict Proca theory, i.e.\ the $\xi\to\infty$ limit (the unitary gauge) of the Stueckelberg theory, while in fact it is only transverse for $\xi \to 0$ (the Landau gauge) as shown explicitly in Sec.~\ref{divergencefeynman}.

In flat space the Feynman propagator for the Stueckelberg theory
\begin{equation}
G_{ab'}^{\mathrm{F}(\xi)}(x-x') = - \mathi \bra{0} \mathcal{T} A_a(x) A_{b'}(x') \ket{0} \eqend{,}
\end{equation}
satisfies the equation
\begin{equation}
\label{stueck}
-2 \partial^b \partial_{[b} G_{a]b'}^{\mathrm{F}(\xi)}(x-x')
+ \xi^{-1} \partial_a \partial^b G_{bb'}^{\mathrm{F}(\xi)}(x-x') + m^2 G_{ab'}^{\mathrm{F}(\xi)}(x-x') = \eta_{ab'} \delta^n(x-x') \eqend{.}
\end{equation}
Hence, the Feynman propagator for the Proca theory, $G^{\mathrm{F}(\infty)}_{ab'}(x-x')$, satisfies this equation with $\xi^{-1}\to 0$. By using the flat-space limit of \eqref{divergenceG} we can write this equation as
\begin{equation}
- \partial^b \partial_b G_{ab'}^{\mathrm{F}(\infty)}(x-x')
+ m^2 G_{ab'}^{\mathrm{F}(\infty)}(x-x') = \left( \eta_{ab'} - \frac{1}{m^2} \partial_a \partial_{b'} \right) \delta^n(x-x') \eqend{,}
\end{equation}
which Tsamis and Woodard find problematic for some reason. (Here we are using the convention $\partial_{b'} = \partial/\partial x^{b'}$ again.) However, this is the consequence of \eqref{stueck} satisfied by the Feynman propagator with arbitrary $\xi$, and there is nothing wrong with it. They correctly point out, however, that the Feynman propagator in the Landau gauge, $G^{\mathrm{F}(0)}_{ab'}(x-x')$, satisfies
\begin{equation}
 - \partial^b \partial_b G_{ab'}^{\mathrm{F}(0)}(x-x')
+ m^2 G_{ab'}^{\mathrm{F}(0)}(x-x') = \eta_{ab'} \delta^n(x-x') - \partial_a \partial_{b'}D_0^{\mathrm{F}}(x-x') \eqend{,}
\end{equation}
where $D_0^{\mathrm{F}}(x-x')$ is the Feynman propagator for the massless scalar field. This equation can readily be understood as the $\xi \to 0$ limit of~\eqref{stueck} as Tsamis and Woodard note in Ref.~\onlinecite{tsamiswoodard2007}.

\section{Conclusion}

We have obtained the two-point function of a massive vector field described by the Stueckelberg Lagrangian parametrized by $\xi$ using the mode-sum method. Such a Lagrangian arises in the electroweak sector of the standard model of particle physics from spontaneous symmetry breaking, with the gauge symmetry fixed by the usual covariant gauge-fixing term with $\xi$ as a gauge parameter~\cite{itzyksonzuber}. We checked our calculations by considering various limits, including the recovery of the Minkowski two-point function as the radius of de~Sitter space goes to infinity. A comparison with previous results revealed that the two-point function obtained by Allen and Jacobson~\cite{allenjacobson1986} corresponds to the Proca theory, while the two-point function obtained by Tsamis and Woodard~\cite{tsamiswoodard2007} is the Stueckelberg propagator in the Landau gauge $\xi \to 0$. Thus, these two two-point functions are simply different limits, $\xi\to\infty$ and $\xi\to0$, of the two-point function in the Stueckelberg theory. In the massless case, the propagator in the Feynman gauge ($\xi = 1$) was obtained by Allen and Jacobson~\cite{allenjacobson1986}, and in the general gauge by Youssef~\cite{youssef2010} for $n=4$. We find agreement with all these results as well.

The propagator obtained in this paper can now be used to study the massive vector field theory in de~Sitter space. Various calculations have been performed in the massless case using the propagators in the Feynman gauge~\cite{kahyawoodard2005,higuchilee2008,higuchileenicholas2009} as well as the Landau gauge~\cite{kahyawoodard2006,prokopectsamiswoodard2006,prokopectsamiswoodard2008a}. Physical results should naturally be independent of $\xi$ (and more generally of any gauge fixing term) if the gauge invariance is correctly incorporated. As an example, one may take the Coleman-Weinberg effective potential~\cite{colemanweinberg1973} for scalar quantum electrodynamics, which has been calculated in the Landau gauge in de Sitter~\cite{allen1983,prokopectsamiswoodard2008b}. This potential is gauge-dependent, and thus may even diverge in exceptional gauges just as the massless vector propagator diverges for $\xi \to \infty$. However, thanks to the Nielsen identities~\cite{nielsen1975,aitchisonfraser1984,johnston1985} the value of the potential at stationary points, which gives physical masses and coupling constants, is independent of any gauge parameter.  We believe that it would be important to check these previous results using the propagator with arbitrary $\xi$. An important point that needs to be considered carefully is the correct treatment of initial and boundary conditions. For example, the naive formula for calculating the retarded response to a source gives a wrong answer~\cite{tsamiswoodard1994,bicakkrtous2001,woodard2004}. When contributions from the initial surface are properly taken into account, the correct response is obtained using the retarded Green's function with any $\xi$~\cite{higuchilee2008,higuchileenicholas2009,facihuguetrenaud2011}. (A related issue are initial state corrections, which can be most easily resolved by taking either an adiabatic vacuum state at past infinity, enforced by an $\mathi \epsilon$ prescription~\cite{froebrouraverdaguer2012,tanakaurakawa2013}, or using the Euclidean/Hartle-Hawking vacuum~\cite{marolfmorrison2010,higuchimarolfmorrison2010,koraitanaka2013}.)

\begin{acknowledgments}
M. F. acknowledges financial support through a FPU scholarship no.~AP2010-5453 including a short stay at an external research institution, as well as partial financial support by the Research Projects MCI FPA2007-66665-C02-02, FPA2010-20807-C02-02, CPAN CSD2007-00042, within the program Consolider-Ingenio 2010, and AGAUR 2009-SGR-00168.  We thank Ph. Spindel for calling our attention to Ref.~\onlinecite{spindelschomblond1976} and the referee for making us aware of Refs.~\onlinecite{allen1983,prokopectsamiswoodard2008b}.
\end{acknowledgments}

\appendix
\section{Evaluation of the integral $I_\nu(Z)$} \label{appA}

In this Appendix, we explicitly calculate the integral $I_\nu(Z)$ using the mode-sum method in $n$ dimensions. This integral is essentially the scalar two-point function. In $n$ dimensions, we obtain for the angular integration
\begin{equation}
\begin{split}
\int f(\abs{\vec{p}}) \mathe^{\mathi \vec{p} \vec{r}} \frac{\total^{n-1} p}{(2\pi)^{n-1}} &= \frac{\Omega_{n-3}}{(2\pi)^{n-1}} \int_0^\infty \int_0^\pi f(q) \mathe^{\mathi q r \cos \theta} \sin^{n-3} \theta \, q^{n-2} \total \theta \total q \\
&= \frac{r^\frac{3-n}{2}}{(2\pi)^\frac{n-1}{2}} \int_0^\infty f(q) \besselj{\frac{n-3}{2}}{qr} q^\frac{n-1}{2} \total q \eqend{,}
\end{split}
\end{equation}
where we set $r = \abs{\vec{r}}$. Here $\Omega_N= 2\pi^{N+\frac{1}{2}}/\Gamma(\frac{N}{2}+1)$ is the area of the unit $N$-sphere.  Hence, the integral defined by~\eqref{i1_def} reads
\begin{equation}
\label{moreuseful}
\begin{split}
I_\nu(\eta,\eta',\vec{r}) &= 2^{n-2} \pi^\frac{n+2}{2} (\eta\eta')^\frac{n-1}{2} \int \hankel1\nu{-\abs{\vec{p}} \eta} \hankel2\nu{-\abs{\vec{p}}\eta'} \mathe^{\mathi \vec{p} \vec{r}} \frac{\total^{n-1} p}{(2\pi)^{n-1}} \\
&= 2^\frac{n-3}{2} \pi^\frac{3}{2} (\eta\eta')^\frac{n-1}{2} r^\frac{3-n}{2} \int_0^\infty \hankel1\nu{-q\eta} \hankel2\nu{-q \eta'} \besselj{\frac{n-3}{2}}{qr} q^\frac{n-1}{2} \total q \eqend{.}
\end{split}
\end{equation}
This integral can be evaluated by expressing the Hankel functions by Bessel functions and using a result due to Bailey~\cite{bailey1936}. This result of Bailey's in the general case is
\begin{equation}
\label{appendix_propagatorintegrals_3besselint_f4}
\begin{split}
&\int_0^\infty \besselj\mu{ax} \besselj\nu{bx} \besselj\rho{cx} x^{\lambda-1} \total x = \frac{2^{\lambda-1}}{\pi c^{\lambda+\mu+\nu}} a^\mu b^\nu \sin\left( \frac{\pi}{2} (-\rho+\lambda+\mu+\nu) \right) \\
&\quad\times \frac{\Gamma\left( \frac{\rho+\lambda+\mu+\nu}{2} \right) \Gamma\left( \frac{-\rho+\lambda+\mu+\nu}{2} \right)}{\Gamma(1+\mu) \Gamma(1+\nu)} \appell4\left[ \frac{\rho+\lambda+\mu+\nu}{2}, \frac{-\rho+\lambda+\mu+\nu}{2}; 1+\mu, 1+\nu; \frac{a^2}{c^2}, \frac{b^2}{c^2} \right] \eqend{,}
\end{split}
\end{equation}
where $\appell4$ is the fourth Appell hypergeometric function defined by
\begin{equation}
\appell4\left( a,b;c,d;x,y \right) = \sum_{k,l=0}^\infty \frac{(a)_{m+n} (b)_{m+n}}{(c)_m (d)_n m! n!} x^m y^n \eqend{.}
\end{equation}
The integral is convergent if $a,b,c>0$ with $c>a+b$ as well as $\Re(\lambda) < \frac{5}{2}$ and $\Re (\rho+\lambda+\mu+\nu) > 0$. In the case where $\lambda=\rho+2$ and $\mu = \pm \nu$, which is all we need for evaluating $I_\nu(Z)$, Eq.~\eqref{appendix_propagatorintegrals_3besselint_f4} simplifies as
\begin{subequations}
\begin{align}
\label{null_integral}
\int_0^\infty \besselj\nu{ax} \besselj{-\nu}{bx} \besselj\rho{cx} x^{\rho+1} \total x &= 0 \eqend{,} \\
\label{secondbessel}
\int_0^\infty \besselj\nu{ax} \besselj\nu{bx} \besselj\rho{cx} x^{\rho+1} \total x &= \frac{2^{\rho+1}}{\pi c^{\rho+2+2\nu}} a^\nu b^\nu \sin\left( \pi (1+\nu) \right) \frac{\Gamma\left( \rho+1+\nu \right)}{\Gamma(1+\nu)} \nonumber \\
&\qquad\times\appell4\left( \rho+1+\nu, 1+\nu; 1+\nu, 1+\nu; \frac{a^2}{c^2}, \frac{b^2}{c^2} \right) \eqend{.}
\end{align}
\end{subequations}

Now we combine the formula
\begin{equation}
\appell4\left( \alpha, \beta; \beta, \beta; - \frac{x}{(1-x)(1-y)}, -\frac{y}{(1-x)(1-y)} \right) = (1-x)^\alpha (1-y)^\alpha \hypergeom{2}{1}\left( \alpha, 1+\alpha-\beta; \beta; xy \right)
\end{equation}
due to Bailey and the transformation formula for the Gauß hypergeometric function
\begin{equation}
\label{appendix_propagatorintegrals_gaussquadratic}
\hypergeom{2}{1}\left( \alpha, 1+\alpha-\beta; \beta; z^2 \right) = (1-z)^{-2\alpha} \hypergeom{2}{1}\left( \alpha, \beta-\frac{1}{2}; 2\beta-1; - \frac{4 z}{(1-z)^2} \right)
\end{equation}
to find
\begin{equation}
\label{Appellappell}
\begin{split}
&\appell4\left( \alpha, \beta; \beta, \beta; - \frac{x}{(1-x)(1-y)}, -\frac{y}{(1-x)(1-y)} \right) \\
&\qquad= (1-x)^\alpha (1-y)^\alpha(1-\sqrt{xy})^{-2\alpha}\hypergeom{2}{1}
\left( \alpha,\beta-\frac{1}{2};2\beta-1;-\frac{4\sqrt{xy}}{(1-\sqrt{xy})^2} \right) \eqend{.}
\end{split}
\end{equation}
Introducing
\begin{equation}
\label{appendix_propagatorintegrals_uv_def1}
u^2 = - \frac{x}{(1-x)(1-y)} \eqend{,} \qquad v^2 = - \frac{y}{(1-x)(1-y)}
\end{equation}
with the (relative) sign defined by
\begin{equation}
\label{appendix_propagatorintegrals_uv_def2}
uv = \frac{\sqrt{xy}}{(1-x)(1-y)} \eqend{,}
\end{equation}
one obtains after some algebra
\begin{equation}
- \frac{4 \sqrt{xy}}{(1-\sqrt{xy})^2} = \frac{4 uv}{(u+v)^2-1} \eqend{,} \qquad \frac{(1-x)(1-y)}{(1-\sqrt{xy})^2} = \frac{1}{1-(u+v)^2} \eqend{.}
\end{equation}
By substituting these formulas in \eqref{Appellappell} we obtain
\begin{equation}
\label{appendix_propagatorintegrals_f4to2f1_b}
\appell4\left( \alpha, \beta; \beta, \beta; u^2, v^2 \right) = \left( 1-(u+v)^2 \right)^{-\alpha} \hypergeom{2}{1}\left( \alpha, \beta-\frac{1}{2}; 2\beta-1; \frac{4 uv}{(u+v)^2-1} \right) \eqend{.}
\end{equation}
By using this formula in \eqref{secondbessel} we find
\begin{equation}
\label{veryuseful}
\begin{split}
&\int_0^\infty \besselj\nu{ax} \besselj\nu{bx} \besselj\rho{cx} x^{\rho+1} \total x = - \frac{2^{\rho+1}}{\pi} \sin\left( \pi \nu \right) \frac{\Gamma\left( \rho+1+\nu \right)}{\Gamma(1+\nu)} \\
&\qquad\times c^\rho a^\nu b^\nu \left( c^2-(a+b)^2 \right)^{-(\rho+1+\nu)} \hypergeom{2}{1}\left( \rho+1+\nu, \frac{1}{2}+\nu; 1+2\nu; \frac{4 a b}{(a+b)^2-c^2} \right) \eqend{.} \\
\end{split}
\end{equation}
We write the Hankel functions in \eqref{moreuseful} in terms of Bessel functions
\begin{equation}
\hankel1\nu{z} = \frac{\besselj{-\nu}{z} - \mathe^{-\mathi \pi \nu} \besselj\nu{z}}{\mathi \sin (\pi \nu)} \eqend{,} \quad \hankel2\nu{z} = \frac{\mathe^{\mathi \pi \nu} \besselj\nu{z} - \besselj{-\nu}{z}}{\mathi \sin (\pi \nu)} \eqend{,}
\end{equation}
use \eqref{null_integral} and \eqref{veryuseful} and apply a standard hypergeometric transformation which sends the argument to its inverse. We also use the duplication and reflection formulas for the $\Gamma$ function
\begin{equation}
\Gamma(\pm 2\nu) = \frac{\Gamma(\pm \nu) \Gamma\left( \frac{1}{2} \pm \nu \right)}{2^{1\mp 2\nu} \sqrt{\pi}} \eqend{,} \qquad \Gamma(\pm \nu) \Gamma(1\mp\nu) = \pm \frac{\pi}{\sin(\pi \nu)} \eqend{.}
\end{equation}
Thus, we obtain
\begin{equation}
\label{appendix_i1_reducible}
\begin{split}
&\int_0^\infty \hankel1\nu{ax} \hankel2\nu{bx} \besselj\rho{cx} x^{\rho+1} \total x = \frac{1}{\pi^\frac{3}{2} ab} \left( \frac{c}{2ab} \right)^\rho \frac{\Gamma(\rho+1+\nu) \Gamma(\rho+1-\nu)}{\Gamma\left( \rho+\frac{3}{2} \right)} \\
&\qquad\times \hypergeom{2}{1}\left( \rho+1+\nu, \rho+1-\nu; \rho+\frac{3}{2}; \frac{(a+b)^2-c^2}{4 a b} \right) \eqend{.}
\end{split}
\end{equation}

This equation was derived under the condition $c > a+b$, and all transformations we have applied were valid under this condition. We also needed the condition $\Re(\rho) < \frac{1}{2}$. (We also need the condition $-1+\abs{\Re(\nu)} < \Re(\rho)$, but it is always satisfied in the cases we are interested in except for the case $m^2=0$ for $\partial_a I_\mu(Z)$, which can be found by taking the limit $m^2\to 0$ of the result for positive mass.) However, now we can use analytic continuation to affirm that this formula is actually valid for all $a$, $b$ and $c$. To this end we recall that the asymptotic expansion of the Hankel functions at large argument is given by
\begin{equation}
\label{hankel_expansion}
\hankel1\nu{ax} \approx \sqrt{\frac{2}{\pi ax}} \mathe^{\mathi[ ax - (2\nu+1) \pi/4 ]} \eqend{,} \qquad \hankel2\nu{bx} \approx \sqrt{\frac{2}{\pi bx}} \mathe^{-\mathi[ bx - (2\nu+1) \pi/4 ]} \eqend{.}
\end{equation}
Hence the integral is convergent if we let $a \to a (1 + \mathi \epsilon)$ and $b \to b (1 - \mathi \epsilon)$ with $\epsilon>0$ for all positive values of $a$, $b$ and $c$ and for all $\rho$ satisfying $-1+\abs{\Re(\nu)} < \Re(\rho)$. This observation determines how the hypergeometric function in \eqref{appendix_i1_reducible} should be continued to $(a+b)^2-c^2 > 4ab > 0$. Thus, the argument of the hypergeometric function is given, in the limit $\epsilon \to 0$, by
\begin{equation}
\frac{(a+b)^2 - c^2}{4 a b} + \mathi \epsilon \sgn(a-b) \eqend{.}
\end{equation}
Hence, by letting $a=-\eta(1+\mathi\epsilon)$, $b=-\eta'(1-\mathi\epsilon)$ and $c=\abs{\vec{r}}$ in \eqref{appendix_i1_reducible} we find indeed
\eqref{i1_z}, i.e.
\begin{equation}
\label{appendix_propagatorintegrals_i1_result}
I_\nu(\eta,\eta',\vec{r}) = \frac{\Gamma\left( \frac{n-1}{2}+\nu \right) \Gamma\left( \frac{n-1}{2}-\nu \right)}{\Gamma\left( \frac{n}{2} \right)} \hypergeom{2}{1}\left[ \frac{n-1}{2}+\nu, \frac{n-1}{2}-\nu; \frac{n}{2}; \frac{1+Z}{2} - \mathi \epsilon \sgn(\eta-\eta') \right]
\end{equation}
with
\begin{equation}
Z = 1 - \frac{\vec{r}^2 - (\eta-\eta')^2}{2\eta\eta'} \eqend{.}
\end{equation}

\section{The spatial component of the two-point function}
\label{appB}
In this Appendix we show that
\begin{equation}
\begin{split}
N_{\alpha\beta} &= m^2 (H^2\eta\eta')^{-1} \left( \eta_{\alpha\beta} - \frac{\partial_\alpha \partial_\beta}{\laplace} \right) I_\nu(\eta,\eta',\vec{x}-\vec{x}') \\
&\quad- (H^2\eta\eta')^{n-2} \frac{\partial_\alpha \partial_{\beta'}}{\laplace} \partial_\eta \partial_{\eta'} \left( (H^2\eta\eta')^{2-n} I_\nu(\eta,\eta',\vec{x}-\vec{x}') \right)
\end{split}
\end{equation}
and
\begin{equation}
\begin{split}
M_{\alpha\beta} &= \left( (n-1) I'_\nu(Z) + Z I''_\nu(Z) \right) \left( \partial_\alpha Z \right) \left( \partial_{\beta'} Z \right) \\
&\quad+ \left( - (n-1) Z I'_\nu(Z) + (1-Z^2) I''_\nu(Z) \right) \partial_\alpha \partial_{\beta'} Z
\end{split}
\end{equation}
are equal. This establishes our claim that \eqref{space-space} is the spatial component of the two-point function given by~\eqref{wightman_result}.

We first show that
\begin{equation}
\label{traceN}
\eta^{\alpha\beta} N_{\alpha\beta} = m^2 (H^2\eta\eta')^{-1} (n-2) I_\nu(Z) + (H^2\eta\eta')^{n-2} \partial_\eta \partial_{\eta'} \left( (H^2\eta\eta')^{2-n} I_\nu(Z) \right)
\end{equation}
and
\begin{equation}
\label{traceM}
\begin{split}
\eta^{\alpha\beta} M_{\alpha\beta} &= - \left( (n-1) I'_\nu(Z) + Z I''_\nu(Z) \right) \frac{\eta^2 + (\eta')^2}{\eta^2(\eta')^2} \\
&\quad+ \left( (3-n) (n-1) Z I'_\nu(Z) + (n-1 + (3-n) Z^2) I''_\nu(Z) \right) \frac{1}{\eta\eta'}
\end{split}
\end{equation}
are equal. To do so, we use the expressions~\eqref{z_time_der} and \eqref{z_spatial_der} for the derivatives of $Z$ and the fact that $I_\nu(Z)$ only depends on $Z$. We first find
\begin{equation}
\begin{split}
&(H^2\eta\eta')^{n-2} \partial_\eta \partial_{\eta'} \left( (H^2\eta\eta')^{2-n} I_\nu(Z) \right) = - \left( (n-1) I'_\nu(Z) + Z I''_\nu(Z) \right) \frac{\eta^2 + (\eta')^2}{\eta^2(\eta')^2} \\
&\qquad+ \left( (n-2)^2 I_\nu(Z) + (2n-3) Z I_\nu'(Z) + (1+Z^2) I_\nu''(Z) \right) \frac{1}{\eta\eta'} \eqend{.}
\end{split}
\end{equation}
By substituting this expression in \eqref{traceN} and using the hypergeometric equation \eqref{simple-hyper} satisfied by $I_\nu(Z)$ we indeed establish that $\eta^{\alpha\beta} M_{\alpha\beta} = \eta^{\alpha\beta} N_{\alpha\beta}$.

Next we prove that the traceless parts of $M_{\alpha\beta}$ and $N_{\alpha\beta}$ also agree. With the notation $\vec{r} = \vec{x}-\vec{x}'$, $r = \abs{\vec{r}}$ the traceless parts can be written as
\begin{equation}
\label{tracelessN}
N'_{\alpha\beta} = (H^2\eta\eta')^{n-2} \left( \frac{\eta_{\alpha\beta}}{n-1} - \frac{\partial_\alpha \partial_\beta}{\laplace} \right) \left( \frac{m^2}{H^2\eta\eta'} - \partial_\eta \partial_{\eta'} \right) \left( (H^2\eta\eta')^{2-n} I_\nu(\eta,\eta',\vec{r}) \right)
\end{equation}
and (using the expressions~\eqref{z_spatial_der})
\begin{equation}
\label{tracelessM}
M'_{\alpha\beta} = - \left( I'_\nu(Z) + \frac{Z}{n-1} I''_\nu(Z) \right) \frac{(n-1) r_\alpha r_\beta - r^2 \eta_{\alpha\beta}}{\eta^2\eta'^2} \eqend{.}
\end{equation}
We use the expression \eqref{moreuseful} for $I_\nu(\eta,\eta',\vec{r})$ in \eqref{tracelessN} and find
\begin{equation}
\label{B8}
\begin{split}
N'_{\alpha\beta} &= 2^\frac{n-3}{2} \pi^\frac{3}{2} \left( \frac{\eta_{\alpha\beta}}{n-1} - \frac{r_\alpha r_\beta}{r^2} \right) (\eta\eta')^\frac{n-3}{2} r^\frac{3-n}{2} \\
&\quad\times \Bigg[ - \frac{m^2}{H^2} \int_0^\infty \hankel1\nu{-q\eta} \hankel2\nu{-q \eta'} \besselj{\frac{n+1}{2}}{qr} q^\frac{n-1}{2} \total q \\
&\qquad+ \int_0^\infty \left( \frac{n-3}{2} \hankel1\nu{-q\eta} + q\eta \hankel[{}']1\nu{-q\eta} \right) \\
&\qquad\quad\times \left( \frac{n-3}{2} \hankel2\nu{-q \eta'} + q\eta' \hankel[{}']2\nu{-q \eta'} \right) \besselj{\frac{n+1}{2}}{qr} q^\frac{n-1}{2} \total q \Bigg] \eqend{.}
\end{split}
\end{equation}
In obtaining this expression, we have used
\begin{equation}
\partial_\alpha \partial_\beta \left[ r^{-\frac{n-3}{2}} \besselj{\frac{n-3}{2}}{qr} \right] = q^2 r^{-\frac{n-3}{2}} \left[ \besselj{\frac{n+1}{2}}{qr} \left( \frac{r_\alpha r_\beta}{r^2} - \frac{\eta_{\alpha\beta}}{n-1} \right) - \frac{\eta_{\alpha\beta}}{n-1} \besselj{\frac{n-3}{2}}{qr} \right] \eqend{,}
\end{equation}
obtained by a straightforward application of the raising and lowering operators for Bessel functions.
Interpreting the derivatives as derivatives with respect to $q$, we can integrate by parts in the second integral in~\eqref{B8}. By using the series expansions of the Bessel and Hankel functions, it can readily be seen that the boundary terms at $q=0$ vanish. Furthermore, by using the $\mathi \epsilon$ prescription introduced after~\eqref{hankel_expansion} we find that the boundary terms at $q=\infty$ vanish as well. Therefore we obtain
\begin{subequations}
\begin{align}
\label{usedhankeleq}
N'_{\alpha\beta} &= 2^\frac{n-3}{2} \pi^\frac{3}{2} \left( \frac{\eta_{\alpha\beta}}{n-1} - \frac{r_\alpha r_\beta}{r^2} \right) (\eta\eta')^\frac{n-3}{2} r^\frac{3-n}{2} \nonumber \\
&\quad\times \Bigg[ r \int_0^\infty \hankel1\nu{-q\eta} \left( \frac{n-3}{2} \hankel2\nu{-q \eta'} + q\eta' \hankel[{}']2\nu{-q \eta'} \right) \besselj{\frac{n-1}{2}}{qr} q^\frac{n+1}{2} \total q \nonumber \\
&\qquad+ (\eta')^2 \int_0^\infty \hankel1\nu{-q\eta} \hankel2\nu{-q \eta'} \besselj{\frac{n+1}{2}}{qr} q^\frac{n+3}{2} \total q \Bigg] \\
\label{further-simplified}
&= 2^\frac{n-3}{2} \pi^\frac{3}{2} \left( \frac{\eta_{\alpha\beta}}{n-1} - \frac{r_\alpha r_\beta}{\vec{r}^2} \right) (\eta\eta')^\frac{n-3}{2} r^\frac{3-n}{2} \nonumber \\
&\quad\times \Bigg[ - r (\eta')^\frac{n-1}{2}\partial_{\eta'} \left( (\eta')^\frac{3-n}{2} \int_0^\infty \hankel1\nu{-q\eta} \hankel2{\nu}{-q\eta'} \besselj{\frac{n-1}{2}}{qr} q^\frac{n+1}{2} \total q \right) \nonumber \\
&\qquad+ (\eta')^2 \int_0^\infty \hankel1\nu{-q\eta} \hankel2\nu{-q \eta'} \besselj{\frac{n+1}{2}}{qr} q^\frac{n+3}{2} \total q \Bigg] \eqend{.}
\end{align}
\end{subequations}
We used the Bessel equation satisfied by $H_\nu^{(2)}(-q\eta')$ to find \eqref{usedhankeleq}. The integrals in \eqref{further-simplified} can be found by \eqref{veryuseful}. Then, the derivative with respect to $\eta'$ can be found using \eqref{z_time_der}. In the end, we obtain
\begin{equation}
\begin{split}
N'_{\alpha\beta} &= \left( r^2 \eta_{\alpha\beta} - (n-1) r_\alpha r_\beta \right) \frac{1}{\eta^2(\eta')^2} \\
&\times\Bigg[ \frac{\Gamma\left( \frac{n+1}{2}+\nu \right) \Gamma\left( \frac{n+1}{2}-\nu \right)}{2 \Gamma\left( \frac{n}{2}+1 \right)} \hypergeom{2}{1}\left( \frac{n+1}{2}+\nu, \frac{n+1}{2}-\nu; \frac{n}{2}+1; \frac{1+Z}{2} \right) \\
&\qquad+ \frac{\Gamma\left( \frac{n+3}{2}+\nu \right) \Gamma\left( \frac{n+3}{2}-\nu \right)}{4 (n-1) \Gamma\left( \frac{n}{2}+2 \right)} Z \hypergeom{2}{1}\left( \frac{n+3}{2}+\nu, \frac{n+3}{2}-\nu; \frac{n}{2}+2; \frac{1+Z}{2} \right) \Bigg] \eqend{,} \\
\end{split}
\end{equation}
which is equal to $M'_{\alpha\beta}$ defined by \eqref{tracelessM}.

\section{The massless limit for $n=2$, $3$ and $4$}
\label{appC}

In this section it is understood that $Z$ and $z$ mean $Z - \mathi \epsilon \sgn(\eta-\eta')$ and $(1+Z)/2 - \mathi \epsilon \sgn(\eta-\eta')$, respectively. To find the massless limit of the vector two-point function for $n = 2$, we note that
\begin{subequations}
\begin{align}
I_\nu(Z) &= \frac{H^2}{m^2} - 1 - \ln(1-z) + \bigo{m^2} \eqend{,} \\
\label{Inu2}
I_\nu'(Z) &= \frac{1}{2} \left( \frac{1}{1-z} + \frac{m^2}{H^2} \frac{\ln(1-z)}{z} \right) + \bigo{m^4} \eqend{.}
\end{align}
\end{subequations}
We also note that $I_\mu'(Z)$ is found by replacing $m^2$ by $\xi m^2$ in \eqref{Inu2} and that $F_\nu(Z) = Z I'_\nu(Z)$. Using these expression in \eqref{AB} and taking the massless limit, we find
\begin{subequations}
\begin{align}
A(Z) &= - 1 - (1+\xi) \frac{\ln(1-z)}{2z} \eqend{,} \\
B(Z) &= - \left( \frac{1}{2} - (1+\xi) \frac{1}{4z} \right) \frac{1}{1-z} + (1+\xi) \frac{\ln(1-z)}{4 z^2} \eqend{.}
\end{align}
\end{subequations}
This result can be shown to agree with that of Allen and Jacobson~\cite{allenjacobson1986} if $\xi = 1$.

The massless two-point function for $n = 3$ was not presented in Ref.~\onlinecite{allenjacobson1986}. From \eqref{I_nu_3D} we readily find
\begin{subequations}
\begin{align}
I_\nu(Z) &= \frac{2}{\sqrt{\pi}} \frac{y}{\sin y} \left( 1 - \frac{m^2}{H^2} \frac{\pi^2 - y^2}{6} \right) + \bigo{m^4} \eqend{,} \\
I_\mu'(Z) &= \frac{2}{\sqrt{\pi}} \left[ - \xi \frac{m^2}{H^2} \frac{y}{2 \sin y} - \frac{1}{\sin y} \frac{\total}{\total y} (y \cot y) \right] + \bigo{m^4} \eqend{,} \\
F_\nu(Z) &= \frac{\total}{\total Z} ( Z I_\nu(Z) ) = - \frac{2}{\sqrt{\pi} \sin y} \frac{\total}{\total y} \left( y \cot y - \frac{1}{6} \frac{m^2}{H^2} ( \pi^2 - y^2) y \cot y \right) + \bigo{m^4} \eqend{.}
\end{align}
\end{subequations}
Then, in the massless limit we obtain
\begin{subequations}
\begin{align}
\lim_{m^2 \to 0} A(Z) &= Q(Z) + \frac{2 y}{\sqrt{\pi} \sin y} \eqend{,} \\
\lim_{m^2 \to 0} B(Z) &= Q'(Z) \eqend{,}
\end{align}
\end{subequations}
where
\begin{equation}
Q(Z) = \frac{1}{\sqrt{\pi}} \left[ \frac{1}{3\sin y} \frac{\total}{\total y} \left( (\pi^2 - y^2) y \cot y \right) + \xi \frac{y}{\sin y} \right] \eqend{.}
\end{equation}

Finally we write down the $n = 4$ case in the Allen-Jacobson notation. We note~\cite{higuchileenicholas2009}
\begin{equation}
\frac{\total}{\total z} \tilde{\Delta}(Z) = \frac{1}{16\pi^2} \left[ \frac{1}{1-z} - \frac{1}{3z} - \frac{2z+1}{3z^2} \ln(1-z) \right] \eqend{.}
\end{equation}
(The factor of $H^2$ is wrong in Ref.~\onlinecite{higuchileenicholas2009}.) Hence
\begin{equation}
\frac{\total^2}{\total z^2} \tilde{\Delta}(Z) = \frac{1}{16\pi^2} \bigg[ \frac{1}{z} + \frac{1}{1-z} + \frac{1}{(1-z)^2} + \frac{2}{3z^2} + \frac{2}{3} \frac{1+z}{z^3} \ln(1-z) \bigg] \eqend{.}
\end{equation}
Then by using
\begin{subequations}
\begin{align}
\partial_a \partial_{b'} Z &= H^2 \left[ g_{ab'} + 2 (1-z) n_a n_{b'} \right] \eqend{,} \\
( \partial_a Z ) ( \partial_{b'} Z ) &= 4 H^2 z (1-z) n_a n_{b'} \eqend{,}
\end{align}
\end{subequations}
we obtain
\begin{equation}
\label{gauge-contr}
\begin{split}
\partial_a \partial_{b'} \tilde{\Delta}(Z) &= \frac{H^2}{16\pi^2} \left[ \frac{1}{2(1-z)} - \frac{1}{6z} - \frac{2z+1}{6z^2} \ln(1-z) \right] g_{ab'} \\
&\quad+ \frac{H^2}{16\pi^2} \left[ \frac{1}{1-z} + \frac{2}{3} + \frac{1}{3z} + \frac{1-z}{3 z^2} \ln(1-z) \right] n_a n_{b'} \eqend{.}
\end{split}
\end{equation}
We also have for $n = 4$
\begin{equation}
\label{non-gauge-contr}
I^{(0)}(Z) = \frac{1}{1-z} \eqend{.}
\end{equation}
By substituting \eqref{gauge-contr} and \eqref{non-gauge-contr} in \eqref{m-less-limit}, we then obtain
\begin{equation}
\lim_{m^2 \to 0} \bra{0} A_a(x) A_{b'}(x') \ket{0} = \alpha(z) g_{ab'} + \beta(z) n_a n_{b'} \eqend{,}
\end{equation}
where
\begin{subequations}
\begin{align}
\alpha(z) &= \frac{H^2}{16\pi^2} \bigg[ \frac{2}{1-z} + (\xi-3) \bigg( \frac{1}{2(1-z)} - \frac{1}{6z} - \frac{2z+1}{6z^2} \ln(1-z) \bigg) \bigg] \\
\beta(z) &= \frac{H^2}{16\pi^2} \bigg[ 2 + \frac{2}{1-z} + (\xi-3) \bigg( \frac{1-z}{3 z^2} \ln(1-z) + \frac{2}{3} + \frac{1}{3z} + \frac{1}{1-z} \bigg) \bigg] \eqend{,}
\end{align}
\end{subequations}
which is the result obtained by Youssef~\cite{youssef2010}.

\bibliography{literature_v3}

\end{document}